  \documentclass[structabstract]{aa}
\usepackage{graphicx}
\usepackage{epstopdf}
\usepackage{txfonts}
\usepackage{natbib}

\bibpunct[]{(}{)}{;}{a}{}{,}

\begin{document}
   \title{Exploring molecular complexity with ALMA (EMoCA): \linebreak Simulations of branched carbon-chain chemistry in Sgr B2(N)}

   \author{
           R.~T. Garrod\inst{1}
           \and
           A. Belloche\inst{2}
           \and
           H.~S.~P. M{\"u}ller\inst{3}
           \and
           K.~M. Menten\inst{2}
           }

   \institute{
              Departments of Chemistry and Astronomy, University of Virginia, Charlottesville, 
              VA 22904, USA\\
              \email{rgarrod@virginia.edu}
              \and
              Max-Planck-Institut f\"ur Radioastronomie, Auf dem H\"ugel 69, 
              53121 Bonn, Germany
              \and
               I.~Physikalisches Institut, Universit{\"a}t zu K{\"o}ln,
              Z{\"u}lpicher Str. 77, 50937 K{\"o}ln, Germany
              }

   \date{Received  / Accepted }

  \abstract
{Using millimeter wavelength data from the Atacama Large Millimeter/submillimeter Array (ALMA), the EMoCA spectral line survey recently revealed the presence of both the straight-chain ({\em normal}) and branched ({\em iso}) forms of propyl cyanide (C$_3$H$_7$CN) toward the Galactic Center star-forming source Sgr~B2(N2). This was the first interstellar detection of a branched aliphatic molecule.}
{Through computational methods, we seek to explain the observed $i$:$n$ ratio for propyl cyanide, and to predict the abundances of the four different forms of the homologous nitrile, butyl cyanide (C$_4$H$_9$CN). We also investigate whether other molecules will show a similar degree of branching, by modeling the chemistry of alkanes up to pentane (C$_5$H$_{12}$).}
{We use the coupled three-phase chemical kinetics model, {\em MAGICKAL}, to simulate the chemistry of the hot-core source Sgr B2(N2), using an updated chemical network that includes grain-surface/ice-mantle formation routes for branched nitriles and alkanes. The network explicitly considers radical species with an unpaired electron on either the primary or secondary carbon in a chain. We also include mechanisms for the addition of the cyanide radical, CN, to hydrocarbons with multiple bonds between carbon atoms, using activation energy barriers from the literature. We use the EMoCA survey data to search for the straight-chain form of butyl cyanide toward Sgr~B2(N2).}
{The observed $i$:$n$ ratio for propyl cyanide is reproduced by the models, with intermediate to fast warm-up timescales providing the most accurate result. Butyl cyanide is predicted to show similar abundances to propyl cyanide, and to exhibit strong branching, with the {\em sec} form clearly dominant over all others. {\em Normal} and {\em iso}-butyl cyanide are expected to have similar abundances to each other, while the {\em tert} form is significantly less abundant. The addition of CN to acetylene and ethene is found to be important to the production of vinyl, ethyl, propyl, and butyl cyanide. The alkanes also show significant branching. We report a non-detection of \textit{n-}C$_4$H$_9$CN toward Sgr~B2(N2), with an abundance at least 1.7 times lower than that of \textit{n-}C$_3$H$_7$CN. This value is within the range predicted by the chemical models.}
{The models indicate that the degree of branching rises with increasing molecular size. The efficiency of CN addition to unsaturated hydrocarbons boosts the abundances of nitriles in the model, and enhances the ratio of straight-to-branched molecule production. Other types of molecule may be less abundant, but show an even greater degree of branching. The predicted abundance of, in particular, $s$-C$_4$H$_9$CN, which at its peak is comparable to that of propyl cyanide, makes it a good candidate for future detection toward Sgr~B2(N2).}


\titlerunning{Simulations of branched carbon-chain chemistry}

\maketitle

%

\section{Introduction}
\label{intro}

Millimeter, sub-millimeter, and centimeter-wavelength spectroscopic observations of high-mass star-forming cores have yielded detections of a range of molecular species, including some of the most complex organic molecules yet detected in the interstellar medium (Herbst \& van Dishoeck 2009). 

The majority of identified organic interstellar molecules are aliphatic, meaning that their carbon atoms are arranged in open, chain-like structures. While aliphatics of three or fewer carbon atoms can necessarily exhibit only a simple straight-chain structure, the presence of four or more carbon atoms allows the possibility of branched carbon chains. The first (and currently, only) interstellar detection of a branched aliphatic molecule came with the discovery by \cite{Belloche14} of {\em iso}-propyl cyanide ($i$-C$_3$H$_7$CN) toward the Galactic Center star-forming source Sagittarius B2(N), as part of the unbiased 3-mm line survey EMoCA (``{\em Exploring molecular complexity with ALMA}''). This survey resolved the emission from two distinct cores within Sgr B2(N), labelled N1 and N2; {\em iso}-propyl cyanide was detected in the more northerly core, N2, along with the straight-chain (or {\em normal}) form of propyl cyanide ($n$-C$_3$H$_7$CN). The latter molecule had already been detected toward SgrB2(N) in an earlier line survey using the IRAM 30~m telescope \citep{Belloche09}. The two isomeric forms of propyl cyanide jointly constitute the largest molecules yet detected toward a star-forming region, so no identifications of larger homologues of either structure have yet been made.

\cite{Belloche14} detected {\em iso}-propyl cyanide in Sgr B2(N2) with a fractional abundance\footnote{Throughout this paper, the abundances of all discussed species are given relative to that of molecular hydrogen.} with respect to H$_2$ of $1.3 \pm 0.2 \times 10^{-8}$, indicating an $i$/$n$ ratio of $0.40 \pm 0.06$. Thus, not only is the branched form of propyl cyanide similarly abundant to other complex organics, but it is very close in abundance to its straight-chain isomer, suggesting that the chemical formation mechanisms of branched molecules may be highly competitive with those of their straight-chain equivalents, for even the smallest molecules capable of such structures.

\cite{Belloche14} also presented chemical kinetics simulations of the coupled gas-phase and dust-grain chemistry occuring in Sgr B2(N), in order to isolate the major formation mechanisms for both forms of propyl cyanide and to explain their abundance ratio. In these models, the majority of complex organic molecules are formed within/upon dust-grain ice mantles through the production and addition of smaller radicals, with the products ultimately desorbing into the gas phase as temperatures rise. These models in fact suggested a bias toward the dominance of the branched structure by a ratio of $i$/$n$ = 2.2. The formation of precursor radicals -- either through the addition of a hydrogen atom to a carbon double-bond or through the abstraction of a hydrogen atom from a saturated carbon chain by a radical such as OH -- was found to favor strongly the production of a radical site at a secondary carbon in a chain, with the subsequent addition of a cyanide or methyl radical then producing $i$-C$_3$H$_7$CN. The only mechanism for propyl cyanide production that was not dominated by the branched form ({\em viz.} CH$_3$CH$_2$ + CH$_2$CN $\rightarrow$ $n$-C$_3$H$_7$CN) involved the addition of radicals that can only exist in the primary (i.e. terminal) form, and thus for which there is no analogous reaction to form branched structures. This route was found to be the only efficient mechanism for the production of the straight-chain form of propyl cyanide.

Two questions naturally arise from the detection of the two isomeric forms of propyl cyanide: firstly, will the ratio of branched to straight-chain forms be preserved or even increased in the case of larger homologues, e.g. butyl cyanide (C$_4$H$_9$CN)? Secondly, are similar ratios to be expected for other, similarly-sized molecules? To begin to answer these questions, here we extend our existing hot-core chemical kinetics model to include a range of reactions and processes dealing with the formation and destruction of butyl cyanide, butane (C$_4$H$_{10}$), and pentane (C$_5$H$_{12}$). We also refine and extend the existing network of reactions for propyl cyanide. Butane, like propyl cyanide, may occur in both {\em normal} and {\em iso} forms. Pentane may exist in three structures ($n$-, $i$-, and $neo$-pentane), while butyl cyanide has four possible structures: {\em normal}, {\em iso} (also known as 3-methylbutyronitrile), {\em sec} (also known as 2-methylbutyronitrile), and {\em tert}. The $n$/$i$/$s$/$t$ nomenclature relates to the the arrangement of the branching methyl group. Each of these structures is illustrated in Fig. 1. Both $t$-C$_4$H$_9$CN and {\em neo}-C$_5$H$_{12}$ may be seen to have two branching methyl groups.

In Sect. 2 we describe the physical and chemical model used. Results are presented in Sect. 3, with discussion in Sect. 4. We present our conclusions in Sect. 5.

\begin{figure}
\centerline{\resizebox{1.0\hsize}{!}{\includegraphics[angle=0]
{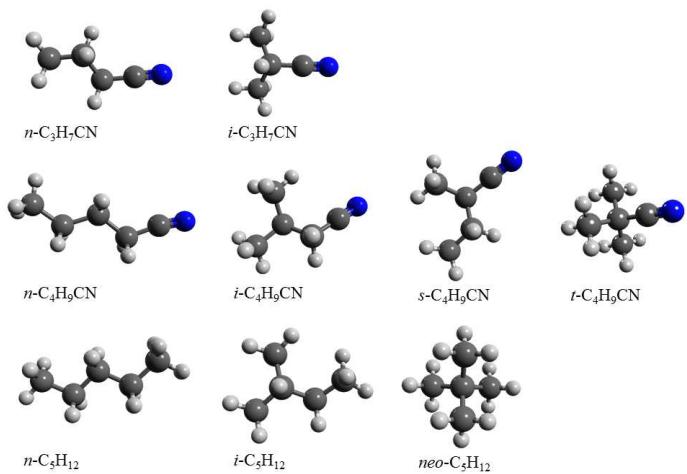}}}
\caption{Structural representations of the straight-chain and branched forms of propyl cyanide, butyl cyanide, and pentane. Only the two forms of propyl cyanide have yet been detected in the interstellar medium. {\em Image credit: E. Willis} }
\label{mol_fig}
\end{figure}

\section {Methods}

The simulations presented here make use of the coupled gas-phase, grain-surface, and ice-mantle chemical kinetics model {\em MAGICKAL}, adopting a set of physical conditions deemed appropriate specifically to the Galactic Center hot-core source Sgr B2(N2). The basic chemical and physical treatments used in the model have been described in detail in past publications; in particular, by \cite{GWWH} and \cite{Garrod13}. A basic outline of the model and a description of any significant alterations are provided below.

\subsection{Physical model}

The physical model consists of two stages: free-fall collapse, followed by a dynamically-static warm-up. A post-collapse gas density of $n_{\mathrm{H}} = 2 \times 10^{8}$ cm$^{-3}$ is assumed, consistent with the density derived by \cite{Belloche14} for the region in core N2 in which the propyl cyanide emission originates. The collapse from 3000 cm$^{-3}$ to the final density occurs over a timescale of $\sim$1 Myr. The increase in visual extinction from 2 to $>$1000 magnitudes during collapse leads to the decrease of dust-grain temperatures from $\sim$18 -- 8 K \citep{GP11}, while gas temperatures are held steady at 10 K throughout the collapse stage. The model uses a single, representative dust-grain radius of 0.1 $\mu$m, with a calculated gas to dust ratio by number of $7.568 \times 10^{11}$.

The warm-up phase follows the treatment of \cite{Garrod13}, with three timescales considered for the warm-up from 8 to 400 K (labeled {\em fast}, {\em medium}, and {\em slow}). During the warm-up, the gas density is assumed to be great enough for the gas and dust temperatures to be well coupled; the dust temperature increases from 8~K until a value equal to the initial gas temperature of 10 K is attained, beyond which both the gas and dust temperatures increase together until the final temperature is reached.

A canonical cosmic-ray ionization rate of $\zeta$ = $1.3 \times 10^{-17}$ s$^{-1}$ is assumed for the dense medium that we model here, in keeping with previous models. The model uses a single, representative dust-grain radius of 0.1 $\mu$m, with a calculated gas to dust ratio by number of $7.568 \times 10^{11}$.

\subsection{Chemical model}

The chemical network is based on that presented by \cite{Belloche14}, and includes the recent additions by \cite{Mueller16}. Initial elemental abundances used in the model are the same as those used by \cite{Garrod13}. The model uses a rate-equation treatment to simulate the gas-phase and ice-mantle chemistry, with the grain-surface chemistry adopting a modified-rate method (Garrod 2008, method ``C''). The atomic and molecular binding energies used in the model, some of which are shown in Table 1, are representative of physisorption to an amorphous water ice surface, following \cite{Garrod13} and earlier models. Chemistry in the outer grain/ice surface (consisting of 1 monolayer or less of ice) is mediated by surface thermal diffusion, with a uniform diffusion barrier ratio of 0.35 with respect to the surface binding energy for each individual species. Reaction within the ice mantle (when present) is assumed to be mediated by a swapping process, with a barrier equal to a fraction 0.7 of the surface binding energy. All reactions and photo-processes occurring in the surface layer are also allowed to occur within the mantle, using the same activation-energy barriers where applicable. Desorption into the gas phase may only occur from the surface layer. Likewise, accretion is allowed to occur only onto the surface layer. 

The surface and mantle abundances are coupled in two ways; firstly, when there is a net addition of material to the surface layer, a corresponding quantity of material is passed into the mantle from the surface, while a net loss from the surface results in the passage of material in the opposite direction. This process thus relates wholly to the overall growth or diminution of the ice (see Garrod 2013 and Hasegawa \& Herbst 1993). 

Secondly, an explicit thermal diffusion between mantle and surface (and vice versa) is allowed, occurring through a similar swapping process as is assumed for reactions within the mantle. The net swapping rate summed over all species must be zero, but the more mobile species will preferentially diffuse outward from the mantle toward the surface, to be replaced by surface atoms or molecules in proportion to their fractional abundances in the surface layer. Garrod (2013, equations 3 \& 4) defined the rates for these complementary processes, to which we here make an adjustment. In order to take full account of the random walk involved in the diffusion to the ice surface of a mantle particle of arbitrary position in the ice, we used a simple Monte Carlo model to determine the relationship between the mean number of discrete diffusion events required for this to occur and the thickness of the ice (see Appendix A). The result of these calculations is a fit to the mean number of ``moves'' required, which is incorporated into the following new expressions that govern the mantle-to-surface and surface-to-mantle exchange (following Garrod 2013):

\begin{equation}
R_{\mathrm{swap},m}(i)=  \frac{1}{2} \, \left(\frac{N_{M}}{N_{S}} + \frac{1}{2} \right)^{-2}     \ k_{\mathrm{swap}}(i) \ N_{m}(i) 
\end{equation}\
\begin{equation}
R_{\mathrm{swap},s}(i)= \frac{N_{s}(i)}{N_{S}} \cdot \sum_{all j} R_{\mathrm{swap},m}(j)
\end{equation}\

\noindent where $R_{\mathrm{swap},m}(i)$ is the rate of mantle-to-surface swapping of mantle species $i$, $R_{\mathrm{swap},s}(i)$ is the rate of surface-to-mantle swapping of surface species $i$, $N_{m}(i)$ and $N_{s}(i)$ are the mantle and surface populations, respectively, of species $i$, and $N_{M}$ and $N_{S}$ are the total mantle and surface populations.

The alteration to the treatment has only a modest effect on the rates of diffusive swapping, due to the relatively thin ices involved in interstellar regimes, typically of order 100 monolayers or less; rates are reduced by a factor on the order of 100 versus the rates assumed by Garrod (2013) at the same dust temperature. This would typically correspond to any particular mantle--surface swapping process becoming important at a temperature on the order of a few K higher than in the Garrod (2013) model. Due to the high barriers involved, no swapping processes or mantle reactions are important around the low temperatures achieved during the collapse phase.

\subsubsection{Tunneling through activation energy barriers.}

Previous implementations of the model have included tunneling through activation energy barriers of reactions occurring on grain surfaces. Following Hasegawa et al. (1992), this involves the calculation of a reaction efficiency based on a rectangular barrier of width 1 \AA, using the reduced mass of the system in the calculation. While such an approach is generally fine for addition reactions, this method underestimates the rates of reactions in which a hydrogen atom is transferred between species (i.e. hydrogen abstraction). In such cases, although neither reacting species is a hydrogen atom, the transferring H atom may undergo tunneling. To address this discrepancy, the code now identifies H-abstraction reactions explicitly, and uses the hydrogen mass instead of the reduced mass for the calculation of the tunneling rate. In all cases, as in previous implementations, the code chooses the faster of the thermal and tunneling rates in the final calculation of the reaction rate.

The reaction rate for a barrier-mediated reaction is dependent not only on the rate of surface diffusion of the reactants, but on a competition between either overcoming the activation energy barrier to reaction, or overcoming the diffusion barrier of one reactant or another (such that the reactants are no longer in contact and cannot react). If diffusion is more rapid than reaction, then the overall reaction rate tends to be low. If tunneling through the reaction barrier is fast, then essentially all encounters between reactants result in a reaction, making the precise value of the reaction efficiency (i.e. tunneling rate) unimportant. The major actor in the abstraction of hydrogen atoms from stable molecules is OH, due to its low barriers to reaction, versus its high barrier to diffusion. As a result, this change makes very little difference in general.

The one regime where the change becomes important is in the relative production rates of different products of the same reaction. The most pertinent such reactions here are those that produce either a primary or a secondary radical as the result of H-abstraction from a carbon chain. As noted by Belloche et al. (2014), barriers to the production of primary radicals are typically higher than those leading to radicals with the radical site at the second carbon in the chain. A purely thermal treatment of the barrier-mediated reaction rates would produce a strong bias toward the reaction with the lower barrier. Using the above-described tunneling treatment leads to a weaker dependence on the barrier height, and thus to a more balanced branching ratio. For example, for two competing H-abstraction reactions with the same reactants but different products, one with a barrier of 500 K and the other with a barrier of 1000 K, using a pure thermal barrier treatment at say 50 K would lead to a branching ratio of around 20,000:1. By allowing the H atom to tunnel, using the treatment above, the ratio would be around 40:1.

\subsubsection{Chemical network}

The new extended chemical network presented here includes a total of 714 unique chemical species, of which the 300 neutrals can exist as either grain-surface,  ice-mantle, or gas-phase species. Charged species are considered in the gas phase only. The network comprises 13,190 gas-phase, grain-surface, and ice-mantle chemical reactions and processes, including accretion (i.e. adsorption) of gas-phase species onto grains, both thermal and non-thermal desorption back into the gas phase, photo-dissociation, and transfer of molecules between the surface ice layer and the mantle. Besides these, there are 1,114 unique grain-surface processes in the network that may be considered true chemical reactions.

Table 1 lists the molecules newly added to the network, as well as some related species that were already present. {\em Normal}-propyl cyanide and its associated radicals were first added to the network by Belloche et al. (2009), while those related to {\em iso}-propyl cyanide were added by Belloche et al. (2014). Species related to C$_4$H$_9$CN, C$_4$H$_{10}$, and C$_5$H$_{12}$ are entirely new to the network.

The addition of new species and reactions follows the approach outlined by Garrod, Widicus Weaver \& Herbst (2008). It is assumed that, in the absence of other evidence, the formation of any new saturated molecule occurs initially on grain surfaces. Such molecules may typically be formed by the addition of radicals that already existed in the network. Where necessary, new radicals are added, each with their own formation and destruction processes both in the gas phase and on the grains. 

Gas-phase and grain-surface/mantle photodissociation processes are included for all new species, with rates guided by previously determined values for similar species in the gas phase; destruction branches typically involve the dissociation of the molecule into a pair of radicals, or a radical and a hydrogen atom. Photodissociation may occur either directly through the action of photons from the external interstellar radiation field, or from the secondary UV field produced by cosmic rays. In the context of the new complex molecules included in the model, which form under high-extinction conditions, it is the latter mechanism that is important.

Only exothermic reactions are included in the new grain chemistry. All radical-radical addition reactions are assumed to be exothermic, whether or not the precise exothermicities are known. In the case of H-abstraction reactions, which may or may not be exothermic, the enthalpy change of the reaction is directly calculated for this purpose. Enthalpies of formation of pertinent species are listed in Table 1. In some cases, these values were not available in the literature, for example where only a value for the {\em primary} radical has been measured. In such cases, the difference in the enthalpy of formation of a similar pair of molecules is used to estimate the missing value (see Table 1).

In Tables 2 and 3 are shown most reactions, new or pre-existing, included in the network for the hydrocarbons and nitriles; activation energy barriers are listed where appropriate. In cases where multiple hydrogenation steps are involved in the production of the pre-existing species, only those with activation energy barriers are shown. 

Many reactions in Tables 2 and 3 involve the addition of a methyl group to a larger radical, and all new hydrocarbons and nitriles have at least one such formation mechanism. Production of those larger radicals can occur in four different ways: (i) the photodissociation of a stable molecule; (ii) the abstraction of a hydrogen atom from a stable molecule (see Table 4); (iii) the addition of a hydrogen atom or CN radical to a carbon double bond (see below); (iv) the addition of CH$_2$ to a smaller radical. Carbene (or methylene), CH$_2$, is assumed to be in the ground (i.e. diradical) state in this model, and its addition to a radical is assumed to produce another, single radical. 

Structural distinctions are only made between isomeric radicals where subsequent reactions in the network would produce different products; for example the primary and secondary radical forms of C$_2$H$_2$CN are not distinguished, as the only reaction in which either isomeric product may participate is the barrierless addition of a hydrogen atom, to form vinyl cyanide.

In the case of butene, which may take multiple isomeric forms, only a single form, 1-butene, is assumed, based on the most kinetically favored products through the full chain of hydrogenation of C$_4$.

\subsubsection{CN addition to carbon multiple bonds}

Crucial to the new grain-surface/mantle chemical network is the inclusion of CN-group addition to carbon double-bonds, in analogy with H-addition. \cite{Gannon2007} studied experimentally the gas-phase kinetics of CN addition to acetylene (C$_2$H$_2$), ethene (C$_2$H$_4$), propene (C$_3$H$_6$), trans-2-butene and iso-butene (C$_4$H$_8$), finding small or negligible activation energy barriers in each case. Similarly to H addition, the production of a radical site at a secondary carbon is energetically favored, thus placing the CN-group on the {\em terminal} carbon, with the formation of a straight carbon chain being finalized with the subsequent addition of an H atom to the radical site. This means that the order in which H and CN are appended to an unsaturated hydrocarbon is critical to whether a straight-chain or branched form is produced.

For inclusion in the new surface network, we assume that the addition of CN to a double-bonded hydrocarbon exclusively forms a single product, which Gannon et al. suggest would be the outcome in the case of collisional de-excitation. We further assume that this product is the secondary radical only. New surface/ice reactions are included for the addition of CN to C$_2$H$_2$ and C$_2$H$_4$, to which we assign the existing 31 K activation energy barrier already included in the gas-phase network (see below), and to C$_3$H$_6$ and C$_4$H$_8$, for each of which a zero barrier is assumed. Higher barriers are also tested for each of these four reactions, corresponding to the assumed values for the addition of H to each hydrocarbon.

The CN reactions with C$_2$H$_2$ and C$_2$H$_4$ were already present in the gas-phase portion of the network, each with an activation energy barrier of 31 K  (following Smith, Herbst \& Chang 2004). For consistency, the reactions with C$_3$H$_6$ and C$_4$H$_8$ are added to the gas-phase network, assuming a zero barrier for each, as with the grain-surface reactions. We fit a temperature-dependent modified Arrhenius rate to each, based on the rates provided by Gannon et al. (2007) at 195 K and 298 K (assuming the trans-2-butene values in the case of butene). These rates are $k($CN+C$_3$H$_6$$) = 3.18 \times 10^{-10} (T/300$ K$)^{-0.05}$ cm$^3$ s$^{-1}$ and $k($CN+C$_4$H$_8$$) = 2.93 \times 10^{-10} (T/300$ K$)^{-0.465}$ cm$^3$ s$^{-1}$. In each case, the gas-phase reaction is assumed to lead exclusively to the abstraction of a hydrogen atom from the hydrocarbon, producing HCN and a radical, which Gannon et al. show to be the most likely outcome based on thermodynamic considerations (due to the absence of a third body that could otherwise stabilize a single reaction product).

The remaining destruction mechanisms added for the gas-phase species are reactions with the most abundant ionic species, namely C$^+$, He$^+$, H$_3$$^+$, H$_3$O$^+$, and HCO$^+$, resulting either in the dissociation of the molecule into multiple fragments (in the case of He$^+$), charge transfer (C$^+$), or proton transfer (molecular ions). In the latter case, only those reactions are included for which the proton affinity of the neutral reactant is greater than the proton affinity of the neutral product. The ion-molecule rates are calculated using the method of Herbst \& Leung (1986), producing values on the order of $10^{-9}$ cm$^3$ s$^{-1}$ with a $T^{-\frac{1}{2}}$ temperature dependence. Recombination of the protonated molecule with an electron typically results in the destruction of the complex into several fragments, except where a hydrogen atom is ejected, leaving the molecular structure intact, which is assumed to occur in 5\% of cases.

\subsubsection{Activation energy barriers for new abstraction reactions}

Table 4 lists the hydrogen-abstraction reactions included for new or relevant species. Where indicated, values are taken from the literature, which are exclusively based on gas-phase processes; where more than one value exists, the one based on multiple measurements extending to the lowest temperature is chosen. For reactions without a literature value, the enthalpy change associated with the reaction is used in an Evans-Polanyi treatment to obtain an activation energy barrier (while endothermic reactions are omitted entirely). This method requires the barriers to be known for two similar reactions of the same radical, with hydrogen abstraction occurring from the same functional group (these reference reactions are not shown in Table 4 -- see Garrod 2013). The Evans-Polanyi method relies on the trend of the activation energy to be inversely proportional to the exothermicity of the reaction among such groups of reactions. For H abstraction by OH, the method tends to break down, as the activation energies of the reference reactions are small, often producing a negative barrier. In those cases, we assume a non-zero value adopted from a similar reaction. In general, the barriers to H abstraction by OH are unimportant for surface/mantle reactions, because the competition between overcoming the reaction barrier and overcoming the diffusion barrier of OH typically favors the reaction. However, in cases where either a primary or secondary radical could be formed, the relative values of each become important in determining the reaction product. As seen in Table 4, for OH reactions, the barrier to secondary-radical production is typically on the order of a few hundred K lower than that of primary-radical production.

Besides adding new abstraction reactions, we also remove two from the previous Garrod (2013) network corresponding to the abstraction of H from methanol by the CH$_3$ radical. The barrier derived from gas-phase experimental data by Tsang (1987) is based on reaction rates at 300 -- 2000 K, and represents a purely empirical fit to the data. A re-evaluation of the rate coefficients using a standard collisional rate of 10$^{-10}$ cm$^{3}$s$^{-1}$ would suggest a significantly higher barrier to either the formation of CH$_3$O or CH$_2$OH (perhaps twice the 3600 or 3480 K, respectively, that were adopted previously). In the Garrod (2013) model, this reaction on grain surfaces was of minimal importance, but here it can become a significant influence on the methanol-related chemistry, due to the modification of the treatment to H-abstraction rates used in the present model (see Sect. 2.2.1). Here we choose to omit these reactions entirely from the network, on the understanding that a more realistic and significantly higher pair of barriers will again minimize their importance to the grain-surface chemistry.

\section{Results}

Table 5 shows the abundances of the major ice-mantle constituents as produced by the chemical model at the end of the collapse stage, along with observational reference values from the literature. Abundances are expressed as a percentage of the water ice value. With the exception of CO, the calculations are generally in reasonably good agreement with observed values, particularly as compared with Sgr A*. Calculated carbon monoxide abundances are around four times higher than observational estimates, although the higher volatility of CO compared to that of other major ice components could indicate that the observed values are more representative of regions in which temperatures are modestly elevated, versus the 8 K dust temperature produced at the end of the modelled collapse phase. This is also in line with reported dust temperatures toward Sgr B2 of 20 -- 28~K (Guzm{\'a}n et al. 2015), which lie on the threshold over which CO is likely to desorb -- around 25~K in these models. \citep[See][ for a further discussion of dust temperatures in Sgr B2]{Belloche16}.

Figures 2 -- 4 show the calculated fractional abundances (with respect to H$_2$) of the nitriles and hydrocarbons of interest, as well as a selection of important or typically-observed hot-core molecules. The three figures show, respectively, the results from the {\em fast}, {\em medium}, and {\em slow} warm-up models. Solid lines indicate gas-phase abundances; dotted lines of the same colour indicate the same species on the grains (combining grain/ice-surface and mantle abundances). The peak gas-phase abundances with respect to H$_2$ of each of these molecules and the temperatures at which they are achieved are given in Table 6 for each of the warm-up models. (Results from the high-$E_{A}$ models are discussed, in Sect. \ref{highEA}).

The gas-phase abundances achieved for the typically-observed hot-core molecules methanol (CH$_3$OH), methyl formate (HCOOCH$_3$) and dimethyl ether (CH$_3$OCH$_3$) are mostly within acceptable observational bounds for each model, except in the case of methyl formate for the {\em slow} warm-up model. As seen in the results of Garrod (2013) and earlier publications, the relatively low temperature at which this molecule is desorbed from grains makes it particularly susceptible to gas-phase ion-molecule destruction, which becomes more significant with greater warm-up timescale. Abundances for all of the molecules are however in broad agreement with the results of Garrod (2013), noting both that the present model uses a higher warm-up stage gas density and that the results presented in Table 6 are given as a fraction of the abundance of H$_2$, rather than the total hydrogen abundance. Again, HCOOCH$_3$ production occurs almost exclusively through the addition of HCO to CH$_3$O on grains at around 25 -- 30 K. CH$_3$OCH$_3$ forms a little earlier (via CH$_3$ + CH$_3$O) and also desorbs earlier; however, its main production mechanism is the gas-phase reaction between methanol and protonated methanol.

\subsection{Simple nitriles: hydrogen cyanide \& methyl cyanide}

As in previous models, almost all HCN present during the warm-up phase is initially found in the ice mantles, having formed during the collapse phase via grain-surface atomic addition reactions. At around 43 K, the HCN on the grains substantially desorbs into the gas phase (see Figs. 2--4, panel a). 

In the {\em medium} and {\em slow} warm-up models, the most rapid gas-phase destruction mechanism for HCN is protonation to HCNH$^+$, followed by dissociative electronic recombination, which can produce CN or HNC, or re-form HCN (1:2:2). The most important destruction mechanism of the nitrile species in this loop comes through the gas-phase reaction of CN with atomic O to produce CO and N. The protonation-initiated destruction of HCN in these two models becomes most important when the gas-phase H$_2$CO (Figs. 2--4, panel a) that is released from the grains at temperatures around 40 K has been destroyed sufficiently that it no longer dominates the destruction of molecular ions (see Garrod et al. 2008, Garrod 2013), which may then attack other molecules, including HCN. 

In the {\em fast} warm-up model, there is insufficient time to complete the gas-phase destruction of H$_2$CO before a range of other molecules is ejected into the gas at around 100 K, reducing the abundances of reactive ions (relative to the longer-timescale models) and allowing HCN to survive largely unscathed through the rest of the model run. All models show an uptick in gas-phase HCN abundances during the warm period ($>$100 K), and this is most pronounced in the 300 -- 400 K regime, resulting ultimately from the increasing efficacy of gas-phase, barrier-mediated H-abstraction from ammonia by H atoms as gas temperatures increase, ultimately driving up atomic N abundances in the gas phase that feed the formation of HCN.

Methyl cyanide (CH$_3$CN) begins the warm-up phase with modest abundances on the grains, deriving (as with HCN) from atomic addition processes during the cold collapse phase. This process is largely described by the sequence CN $\rightarrow$ C$_2$N $\rightarrow$ HC$_2$N $\rightarrow$ CH$_2$CN $\rightarrow$ CH$_3$CN. In the {\em fast} and {\em medium} warm-up models, the grain-surface abundance of CH$_3$CN increases significantly from around 43 K onward, as the injection of HCN into the gas phase leads to reaction with CH$_3$$^+$ ions, producing protonated methyl cyanide. Following recombination with electrons, the methyl cyanide produced adsorbs onto the surfaces of the dust-grain ice-mantles, remaining there until it desorbs at around 90 K.

In the {\em slow} model, the low and brief elevation of H$_2$CO abundances in the gas phase around 40 K is not sufficient to bring ionic abundances down appreciably compared with the other models. As a result, the reaction between C$^+$ and CH$_4$ produces large quantities of C$_2$H$_3$$^+$ and C$_2$H$_2$$^+$. Dissociative electronic recombination of these ions boosts the abundance of C$_2$H, which reacts with gas-phase atomic N to form C$_2$N. The C$_2$N is rapidly hydrogenated on grains to form CH$_3$CN, whose abundance is strongly enhanced at around 40~K as a result.

In the {\em medium} and {\em slow} warm-up models, abstraction of H from CH$_3$CN by various radicals has sufficient time to bring down the methyl cyanide abundance on the grains significantly, prior to release into the gas phase around 90~K. The resultant CH$_2$CN reacts to form larger nitriles (such as {\em iso}-butyl cyanide).

Beyond around 90--100~K (following desorption from the grains), the gas-phase methyl cyanide abundance grows strongly, mainly as a product of the dissociative recombination of protonated ethyl cyanide, for which a 5\% branching fraction is assumed. As the abundance of ethyl cyanide falls, the radiative association between CH$_3$$^+$ and HCN again becomes the dominant formation route for CH$_3$CN.

\subsection{Vinyl \& ethyl cyanide}

Vinyl cyanide (C$_2$H$_3$CN) begins to form in small amounts on the dust-grain surfaces during the collapse stage, through the hydrogenation of HC$_3$N; however, the hydrogenation continues all the way to ethyl cyanide (C$_2$H$_5$CN), keeping vinyl cyanide abundances low ($\sim$$10^{-12}$) and allowing ethyl cyanide to maintain a steady abundance of a few $\times 10^{-10}$ on the grains, through the collapse phase and on into the warm-up phase. 

The abundance of vinyl cyanide on dust grains becomes more significant (a few $\times 10^{-9} n[$H$_2$]) during the warm-up stage, at around 40 -- 60 K, through mechanisms that rely on both gas-phase and grain-surface processes. The desorption of HCN from the grains at around 40~K results in the production of gas-phase CN (see Sect. 3.1). While this CN may react with gas-phase C$_2$H$_4$ to produce C$_2$H$_3$CN + H (of which the vinyl cyanide may be accreted onto the grain surfaces), the majority of the CN is accreted directly onto the dust grains. The gas-phase reaction between CN and C$_2$H$_2$ produces HC$_3$N, which also accretes onto the grains, where it is hydrogenated to vinyl cyanide (see Table 3, reactions 20 \& 22) and ethyl cyanide. 

The accretion of gas phase-produced CN onto the grains allows a surface reaction (\#21) with C$_2$H$_2$ that produces C$_2$H$_2$CN, which is then hydrogenated to vinyl cyanide. This process is by far the favored channel for vinyl cyanide production on the grains.

The presence of both C$_2$H$_2$ and C$_2$H$_4$ in the gas phase is the result of chemistry associated with the large quantities of methane initially stored in the ices being ejected into the gas at around 25 -- 30~K. The gas-phase C$_2$H$_2$ and C$_2$H$_4$ are both able to accrete onto the grains, and, from there, some is incorporated into the deeper ice mantles until it becomes mobile, returning to the surface at around 55--60~K.

The grain-surface C$_2$H$_3$CN is capable of thermally desorbing at around 100 K, but most of it is destroyed on the grains before this time, being hydrogenated to ethyl cyanide. Vinyl cyanide only achieves an appreciable abundance in the gas phase beyond 100 K, at which time any gas phase-produced C$_2$H$_3$CN that sticks to the grain surfaces is rapidly desorbed before it can be destroyed. It is formed in the gas phase as a by-product of the destruction of ethyl cyanide via protonation and dissociative recombination; the assumed efficiency of vinyl cyanide formation from recombination of protonated ethyl cyanide in the model is 40\%, approximately in line with the experimental values of Vigren et al. (2010).

At a temperature around 30 K, the ethyl cyanide abundance on the grains begins to rise modestly, due to the addition of CN to C$_2$H$_4$ on the grains (reaction 27), followed by hydrogenation (reaction 29). (The equivalent reaction in the gas-phase instead forms vinyl cyanide and a hydrogen atom). The rising abundance of vinyl cyanide that begins around 40 K, mentioned above, produces large quantities of ethyl cyanide via hydrogenation. The rapid conversion of vinyl cyanide to ethyl cyanide ensures that the grain-surface abundance of the latter dominates those of the former at all times.

In the {\em medium} warm-up timescale model in particular, a later bump (at $\sim$65~K) appears in the abundance of grain-surface ethyl cyanide, related to an increase in gas-phase CN production. The CN reacts with C$_2$H$_2$ to produce HC$_3$N, which leads to grain-surface vinyl and thence ethyl cyanide formation as described above. The boost in CN abundance is related to the successful removal of formaldehyde from the gas-phase, via ion-molecule reactions, while HCN still retains modest abundances, allowing the reaction HCO$^+$ + HCN $\rightarrow$ H$_2$CN$^+$ + CO to become important as the HCO$^+$ abundance rises. Electronic recombination of H$_2$CN$^+$ produces the CN.

The gas-phase reaction CN+C$_2$H$_4$$\rightarrow$C$_2$H$_3$CN+H is the major gas-phase formation mechanism for vinyl cyanide at temperatures up to around 100~K, but it never produces a gas-phase fractional abundance greater than a few 10$^{-11}$.

\subsection{Propyl cyanide}

The modeled abundances of the two isomeric forms of propyl cyanide are shown in Figs. 2 -- 4, panel (b). At around 25~K,  small quantities of both molecules are formed on grains, through the direct addition of CH$_3$ to the two isomeric C$_2$H$_4$CN radicals. Grain-surface abundances of the two forms of propyl cyanide diverge around 30~K, with $n$-C$_3$H$_7$CN becoming more abundant due to the addition of CN to propylene (C$_3$H$_6$), followed by hydrogenation of the resulting radical.

In the 40 -- 60~K range, again the reaction of CH$_3$ with the two C$_2$H$_4$CN radicals, formed as intermediates in the production of vinyl and ethyl cyanide (which becomes rapid at these temperatures -- see Sect. 3.2), leads to increased production of {\em normal} and {\em iso}-propyl cyanide. At this temperature range in particular, $i$-C$_3$H$_7$CN production is kinetically favored, due to the rapidity of the CN + C$_2$H$_2$ route to vinyl cyanide and thence to ethyl cyanide. The hydrogenation of vinyl cyanide strongly favors the production of a radical site on the secondary carbon atom, so that the addition of a methyl group to this molecule produces $i$-C$_3$H$_7$CN. The less productive CN + C$_2$H$_4$ $\rightarrow$ $\overset{\centerdot}{\mathrm{C}}$H$_2$CH$_2$CN
process, which bypasses vinyl cyanide, produces a radical site on the primary carbon atom, and is responsible for the production of the {\em normal} form, in this temperature range.

Beyond 60~K, the main formation mechanisms of the two forms of propyl cyanide vary between the different warm-up timescale models. In the case of the {\em medium} warm-up timescale model, the CN + C$_3$H$_6$ reaction contributes very strongly to $n$-C$_3$H$_7$CN formation on grains, as a result of the boost in gas-phase CN abundance described in Sect. 3.2, some of which is accreted onto grains where it may react or otherwise desorb back into the gas phase. Most of this contribution comes at temperatures around 65 -- 72~K. Beyond this temperature, the reaction between C$_2$H$_5$ and CH$_2$CN takes over as the dominant route to $n$-C$_3$H$_7$CN. This pathway is also enhanced by the elevated CN abundance, as H-abstraction from C$_2$H$_6$ by the CN radical (reaction 31) becomes the dominant route to C$_2$H$_5$ formation. The larger grain-surface CH$_3$CN abundance (which is related to the high abundance of CN as described in Sect. 3.1) also contributes, as it is the primary source of CH$_2$CN via H-abstraction by OH on the grains.

While there is a boost in grain-surface $i$-C$_3$H$_7$CN abundance in the {\em medium} warm-up model (associated with the high CN abundance), it is less pronounced than for $n$-C$_3$H$_7$CN, and is insufficient to keep the $i$:$n$ ratio greater than unity. This dominance of the straight-chain form continues into the gas phase as the two forms desorb from the grains at around 160~K. Formation of $i$-C$_3$H$_7$CN occurs primarily through the addition of CH$_3$ to the  CH$_3$$\overset{\centerdot}{\mathrm{C}}$HCN 
radical produced by the hydrogenation of vinyl cyanide.

In the {\em fast} model, the above-described boost in CN abundances in the gas phase and on the grains (caused by the fall in gas-phase formaldehyde abundance) does not occur, so there is no commensurate boost that favors $n$-C$_3$H$_7$CN production over the {\em iso} form. Here, most production of C$_3$H$_7$CN occurs in the 80 -- 110~K range, with $i$-C$_3$H$_7$CN again forming as a result of methyl-group addition to radicals formed through hydrogenation of vinyl cyanide. The onset specifically at $\sim$80~K is related to the release of vinyl cyanide from the deeper grain mantles onto the ice surface, prior  to desorption. $n$-C$_3$H$_7$CN production comes via ethyl-group addition to CH$_2$CN, but in smaller quantities due to the lower quantities of C$_2$H$_5$ available caused by weaker H-abstraction from ethane by CN radicals. Around 110~K, the addition of CN to C$_3$H$_6$ takes over as the dominant process, accounting for $\sim$30\% of the final $n$-C$_3$H$_7$CN abundance on the grains.

In the {\em slow} model, the elevated CN abundance associated with gas-phase HCN destruction does not extend to long times/high temperatures, meaning that the enhanced production of {\em normal} propyl cyanide is not seen here either. H-abstraction from ethyl cyanide followed by methyl-group addition is the major production route for both forms of propyl cyanide in this model, which again favors production of the {\em iso} form.

\subsection{Butyl cyanide}

The most abundant forms of butyl cyanide produced in the models are found in all cases to reach peak gas-phase abundances of the same order of magnitude as the propyl cyanides in those same models. However, it is noteworthy that {\em normal} butyl cyanide is never the dominant form in the gas phase, although until a temperature of around 50~K is reached, it is indeed the dominant form on the grains. The {\em sec} form is the dominant gas-phase form in all models, although in the {\em slow} warm-up model, the {\em iso} form is also competitive. The doubly-branched {\em tert} form is consistently the least abundant form of butyl cyanide in all models.

The initial formation of the {\em normal} and {\em sec} forms of C$_4$H$_9$CN on grains begins around 25~K, through the addition of CN to one of the straight-chain isomeric forms of C$_4$H$_9$. Following the release of grain-surface methane into the gas at this temperature, gas-phase reactions between hydrocarbon ions and CH$_4$ lead to the formation of C$_4$H$_2$, which sticks to and is hydrogenated on the icy grain surfaces. At these temperatures, the main source of CN is the cosmic ray-induced photodissociation of the large amount of HCN on the grains. Production of $i$-C$_4$H$_9$CN and $t$-C$_4$H$_9$CN, on the other hand, requires the formation of a branched C$_4$H$_9$ radical, either through the addition of CH$_2$ to a C$_3$H$_7$ secondary radical or by the photodissociation of $i$-C$_4$H$_{10}$.

At around 30~K, $n$-C$_4$H$_9$CN production on the grains begins to dominate, as the insertion of CN into C$_4$H$_8$ (reaction 50) becomes important. The increased CN abundance at around 40~K due to HCN desorption and gas-phase destruction further increases the influence of this process, bringing the abundance of the $n$- form to around 10$^{-11}$ $n$[H$_2$] on the grains.

In the {\em medium} warm-up model, all forms of butyl cyanide undergo a period of rapid formation at around 55--60~K, for similar reasons as in the case of C$_3$H$_7$CN. The production of the dominant $s$- and $n$- forms occurs through the addition of the C$_2$H$_5$ radical to either the secondary- or primary-radical form, respectively, of C$_2$H$_4$CN, which are produced as described in Sect. 3.2, with the secondary-radical form dominating. By around 75~K, $s$-C$_4$H$_9$CN has become the dominant form. For {\em normal} butyl cyanide, the addition of the C$_3$H$_7$ radical to CH$_2$CN accounts for a few percent of its production; this mechanism therefore does not dominate, unlike the equivalent process for {\em normal} propyl cyanide production (i.e. C$_2$H$_5$ + CH$_2$CN, reaction 34). $i$-C$_4$H$_9$CN is a little less abundant than the {\em normal} form of butyl cyanide, and is produced strongly at around 60~K by the CN + C$_3$H$_6$ reaction (followed by methyl-group addition), with production from 80~K onward occurring through the addition of the secondary-radical form of C$_3$H$_7$ to CH$_2$CN (reaction 53). The {\em tert} form of butyl cyanide requires H-abstraction from $i$-C$_3$H$_7$CN followed by the addition of a methyl group to the resulting radical, whose radical site lies at the centrally-positioned carbon. 

In the {\em fast} model, both the {\em normal} and {\em sec} forms are produced rapidly around 55~K, through the addition of C$_2$H$_5$ to the C$_2$H$_4$CN radicals, but without the much stronger boost at 60~K that occurs for the {\em medium} warm-up model. The sharp growth in both molecules is caused by the migration of C$_2$H$_2$ and C$_2$H$_4$ from the ice mantles to the surfaces, where reaction with CN occurs. $s$-C$_4$H$_9$CN becomes dominant over $n$-C$_4$H$_9$CN on the grains around 75~K, as the hydrogenation of vinyl cyanide -- predominantly forming the secondary radical -- becomes more important. Hydrogenation of propylene, followed by the addition of CH$_2$CN to the resultant radical, is the major formation mechanism for $i$-C$_4$H$_9$CN, as in the {\em medium} timescale model. A final boost to the $s$ form is received up until around 100~K, due to the hydrogenation of vinyl cyanide and the more effective production of C$_2$H$_5$ through H-abtraction from ethane by OH.

In the {\em slow} model, production rates of the $n$, $s$, and $i$ forms of butyl cyanide are more closely matched (versus the other models), peaking around 80~K, although the {\em tert} form is still around two orders of magnitude less abundant than the others. The greater production of $i$-C$_4$H$_9$CN in this model is the result of the more complete destruction of CH$_3$CN via H-abstraction, allowing more significant reaction between CH$_2$CN and the secondary C$_3$H$_7$ radical. 

The desorption of all forms of butyl cyanide occurs around 200~K in all models.

\subsection{Alkanes}

Figures 2 -- 4, panel (d), show the time-dependent abundances of the two forms of butane and the three forms of pentane. The chemical model includes reactions related to the cold-cloud gas-phase chemistry of straight-chain C$_4$ molecules, as well as grain-surface hydrogenation processes all the way to C$_4$H$_{10}$. As a result, a modest abundance of $n$-C$_4$H$_{10}$ (a few $\times 10^{-10} n[$H$_2$]) is built up on the grains during the collapse phase. Approaching a temperature of 20~K in the warm-up phase, $i$-C$_4$H$_{10}$ is formed in the ice mantles through the addition of a methyl group to the secondary radical form of C$_3$H$_7$, which is produced by hydrogenation of propylene. A small amount of $n$-C$_4$H$_{10}$ is also formed through the analogous reaction involving the primary-radical form of C$_3$H$_7$. At these low temperatures, pentane also begins to be formed on the dust grains; cosmic ray-induced photodissociation of either form of C$_4$H$_{10}$ can produce the required radicals, to which a methyl group is then added. The hydrogenation of C$_4$H$_8$, favoring the production of the secondary-radical form of C$_4$H$_9$, allows the {\em iso} form to be produced in slightly higher abundance than the {\em normal} form. Production of {\em neo}-pentane requires the dissociation of $i$-C$_4$H$_{10}$, rather than the more abundant smaller alkanes, leaving it less abundant than its structural isomers. The position of {\em neo}-pentane as the least abundant form and {\em iso}-pentane as the most abundant form is preserved at all temperatures, throughout each of the three models, albeit in varying ratios.

At around 30~K, the abundance of $n$-C$_4$H$_{10}$ receives a large boost, through the gas-phase production of straight-chain C$_4$-related hydrocarbons (primarily C$_4$H$_2$), which are hydrogenated on the grain surfaces. However, the majority of the formation of butane and pentane, for all models, occurs in the $\sim$60--90~K range. {\em Normal} butane is formed predominantly via the addition of C$_2$H$_5$ radicals, starting around 70~K. The {\em iso} form is again produced through the addition of CH$_3$ to the secondary C$_3$H$_7$ radical. {\em Normal}- and {\em iso}-pentane are each formed on the grain surfaces through the addition of a methyl group to the primary or secondary form of the C$_4$H$_9$ radical. The secondary form 
(CH$_3$$\overset{\centerdot}{\mathrm{C}}$HCH$_2$CH$_3$) 
is produced largely through the abstraction of H from $n$-butane by the OH radical, or through the addition of H to C$_4$H$_8$, which continues to be produced through a combination of gas-phase chemistry and grain-surface hydrogenation. The formation of the primary C$_4$H$_9$ radical occurs much more slowly, and only through the photodissociation of $n$-butane. The abundance of $n$-butane and $n$-pentane are therefore closely linked.

{\em Neo}-pentane production at these higher temperatures still occurs through the addition of CH$_3$ to the branched secondary radical form of C$_4$H$_9$, as is the case at lower temperatures. The abundance of {\em neo}-pentane is therefore closely linked to the abundance of {\em iso}-butane. However, unlike {\em normal}-butane, at these higher temperatures at which radicals are more mobile, its precursor radical may be efficiently formed by the abstraction of an H atom from $i$-C$_4$H$_{10}$ by NH$_2$, which occurs much more rapidly than photodissociation. The peak ratio of {\em neo}-pentane to $i$-butane is therefore larger than the ratio of $n$-pentane to $n$-butane.

\subsection{High activation-energy models}

\label{highEA}

In the chemical models described above, low activation energy barriers for reactions of CN with double-bonded hydrocarbons were used (see Table 3). These values, based on the rate data and/or barrier estimates of Gannon et al. (2007), are lower than those for the equivalent reactions involving atomic H. In order to test the dependence of the model results on these barriers, we re-ran the collapse model and the three subsequent warm-up models using barriers to reactions 21, 27, 38, and 50 that we raised to the values assumed for hydrogen addition. 

Results are shown in Table 6, along with those from the low-barrier models. It may be seen that the peak abundance values for molecules not associated with CN are essentially unchanged. Between different warm-up models, the abundances of $n$- and $i$-propyl cyanide and $t$-butyl cyanide are most strongly affected in the {\em fast} model, while the abundances of vinyl cyanide and $n$-, $i$-, and $s$-butyl cyanide are most strongly affected in the {\em slow} model. Ethyl cyanide varies similarly (in factor) between the {\em fast} and {\em slow} models. The {\em medium} warm-up timescale results show only minor differences. The other species are affected by less than $\sim$10\% in all models when using the high activation energy barriers.

Figure 5 plots the chemical abundances calculated for the {\em fast} warm-up model using high CN-related activation energy barriers. The most notable change in these results is the inversion of the $n$:$i$ ratio for propyl cyanide; the {\em normal} form is now dominant. This change is related to the formation of vinyl and ethyl cyanide, as described in Sects. 3.2 and 3.3. In the {\em low-barrier} case, over the temperature range 40--60~K, most vinyl cyanide is formed through the addition of CN to C$_2$H$_2$, followed by reaction with a hydrogen atom. Further hydrogenation of C$_2$H$_3$CN (reactions 24 \& 25) typically produces a secondary radical, allowing {\em iso}-propyl cyanide to form via the addition of a methyl radical. However, in the {\em high-barrier} case, hydrogenation of C$_2$H$_2$ is much more likely to occur than CN addition, resulting in the production of C$_2$H$_3$ and thence C$_2$H$_4$ (and by-passing the production of vinyl cyanide as an intermediate to ethyl cyanide and larger nitriles). This C$_2$H$_4$ may still react with CN at a faster rate than it does with atomic hydrogen (even though more slowly than is possible in the low-barrier case), producing the primary radical 
$\overset{\centerdot}{\mathrm{C}}$H$_2$CH$_2$CN,
from which {\em normal}-propyl cyanide may form. Thus, the ratio in the activation energy barriers (if any) of reactions 21 and 27 plays a significant role in determining the $n$:$i$ ratio for propyl cyanide.

Notable also in the high-barrier, {\em fast} warm-up model is the inversion of the $n$:$i$ ratio for {\em butyl} cyanide, with the {\em iso} form becoming more dominant (while {\em sec} is still the most abundant of all). In fact, {\em iso}-butyl cyanide rises a little, while the $n$ and $s$ forms fall. The rise in $i$-C$_4$H$_9$CN is caused by the larger barrier to the addition of CN to C$_3$H$_6$, which makes hydrogenation of propylene more likely, favoring the production of the secondary radical C$_3$H$_7$, and thence $i$-C$_4$H$_9$CN (via reaction 53). The $n$ and $s$ forms fall a little due to the lower production of both forms of the C$_2$H$_4$CN radical, resulting from the higher barriers.

\subsection{Comparison with observations}

Table 7 shows the ratios of the peak abundances of various nitriles and alkanes, along with observed ratios where available. The $i$:$n$ ratio for propyl cyanide is the correct side of unity for the {\em medium} warm-up timescale models using either the low or high activation energy values for CN addition reactions. The ratio achieved for the {\em medium} warm-up model with low barriers is still within a factor of 2 of the observational value of $0.40 \pm 0.06$ obtained by Belloche et al. (2014). In the {\em fast} and {\em slow} models, this ratio is greater than 1 except for the {\em fast} model with high barriers. The ratio achieved in this case is in agreement  with the observational value, and well within the stated errors.

Since none of the forms of butyl cyanide have yet been detected in the interstellar medium, no ratios exist between the different forms. However, the models show that the ratio of $s$:$n$ is around a factor of 2 or more for all conditions tested. The spread in ratios of $i$:$n$ for each model suggests that the $i$ and $n$ forms are likely to be of similar abundance, while {\em tert}-butyl cyanide is no more than 10\% of the abundance of the {\em normal} form, and sometimes considerably lower.

The ratio of $i$:$n$ for butane is generally smaller than that for propyl cyanide, and is consistently less than unity. This is in sharp contrast to the $i$:$n$ ratio for pentane, which is dominated by the {\em iso} form. {\em Neo}-pentane is consistently less abundant than either of the other forms, but its ratio with {\em normal}-pentane is not as low as the ratio between {\em tert} and {\em normal} butyl cyanide. 

At the bottom of the Table are shown model ratios between homologues of observed straight-chain nitriles, as well as the ratio between $n$-butyl and $n$-propyl cyanide. For this latter ratio, the models suggest a value of around 0.1--1. Based on the spread of values, {\em normal}-butyl cyanide could take an abundance around 50\% that of {\em normal}-propyl cyanide.

Using the spectroscopic predictions of \citet{Ordu12} that are 
available in the CDMS catalog, we searched for \textit{n-}C$_4$H$_9$CN  in the 
EMoCA spectrum of Sgr~B2(N2). Our chemical models predict that the peak of
butyl cyanide abundance in the gas phase should be located at a temperature of 
$\sim$200~K (Table~\ref{tab-abuns}). With a temperature scaling as r$^{-0.83}$ 
\citep[][]{Rolffs11} and a temperature of 150~K for a diameter of 1.0$\arcsec$ 
as traced with propyl cyanide \citep[][]{Belloche14}, the region of Sgr~B2(N2) 
with $T > 200$~K is expected to have a diameter of $< 0.7\arcsec$. Assuming
a source size of 0.7$\arcsec$, a temperature of 200~K, and the same linewidth
as for \textit{n-}C$_3$H$_7$CN, we derive an upper limit to the column density 
of \textit{n-}C$_4$H$_9$CN of $1.2 \times 10^{17}$~cm$^{-2}$ based on the 
\textit{anti-anti} conformer (and after correction of the partition function 
for the conformational (factor 2.9) and vibrational (factor 10.3) 
contributions at 200~K that are not included in the CDMS partition function). 
This column-density upper limit is 1.5 times lower than the column density of 
\textit{n-}C$_3$H$_7$CN toward Sgr~B2(N2) \citep[][]{Belloche14}, which 
translates into a factor 1.7 in terms of abundance relative to H$_2$ if the 
H$_2$ column density is inversely proportional to the radius. The resulting upper limit 
on the abundance ratio [\textit{n-}C$_4$H$_9$CN] / [\textit{n-}C$_3$H$_7$CN] = 0.59
is consistent with the predictions of our chemical models (i.e. 0.1--1). The {\em medium} 
warm-up timescale model predicts an abundance ratio around 0.3, i.e. a factor of 2 lower than can be 
confirmed using the existing EMoCA data.

The models produce ratios of C$_2$H$_5$CN:CH$_3$CN and C$_2$H$_5$CN:C$_2$H$_3$CN that are reasonably close to the observed ratios, based on the spectral fits produced by Belloche et al. (2016); the former ratio is no worse than a factor 6 away for any model, and the latter is no worse than a factor 4. The model ratio between $n$-propyl cyanide and ethyl cyanide is generally larger than the observed value, although the {\em slow} and {\em fast} models with low CN-related barriers come within a factor of 2--3.


\section{Discussion}
\label{discussion}

Whereas the previous model of Belloche et al. (2014) has shown difficulty in reproducing the observed $i$:$n$ ratio for propyl cyanide of $0.40 \pm 0.06$, the models presented here can produce values ranging between around seven times larger to three times smaller than that ratio, depending on both the warm-up timescale and the barriers chosen for CN-related reactions with multiple-bonded carbon chains. While all the propyl cyanide $n$:$i$ ratios produced by the models are therefore on the order of unity, the precise value achieved (and thus the dominance of one form over the other) is not determined by a single process but rather by the interplay between multiple mechanisms. 

In this more complete model of the chemistry associated with propyl cyanide, the main formation mechanism of the straight-chain form is still, in general, the addition of the C$_2$H$_5$ and CH$_2$CN radicals in/upon dust-grain ice mantles, as found by Belloche et al. (2014). In the {\em medium} and {\em fast} timescale warm-up models, a large contributor is also the newly-added reaction of CN with propylene (C$_3$H$_6$), in which the CN radical inserts itself into the carbon double bond, preferentially forming a radical site at the secondary carbon atom. The hydrogenation of this radical then produces $n$-C$_3$H$_7$CN. The {\em slow} warm-up model produces $n$-C$_3$H$_7$CN through the addition of a methyl group to the primary radical produced via H-abstraction from ethyl cyanide.

It is notable that the largest quantities of all nitriles are produced in the {\em medium} warm-up models, and their large abundances are related to the survival of HCN in the gas-phase up to temperatures around 70~K (following release from the grains at $\sim$40~K), by which point most of the nitriles have formed efficiently on grains. The gradual destruction of HCN in the gas phase over this $\sim$40--70~K temperature range produces the CN that then accretes onto the grains to react with radicals and with the most abundant non-radical species. The models with longer or shorter warm-up timescales do not produce this sustained, but moderate, gas-phase conversion of HCN to CN. The survival of HCN is also related to the availability of H$_2$CO in the gas phase, which acts to protect HCN (and other molecules) from rapid ion-molecule destruction, until H$_2$CO itself is destroyed. The particular abundance and observability of nitriles toward Sgr B2(N) may therefore indicate that this set of conditions indeed pertains, and that the {\em medium} warm-up timescale model is the most appropriate for this source; indeed, this model is in good agreement with the observed abundances on the order of 10$^{-8}$ relative to H$_2$ determined by \citet{Belloche14} for both forms of propyl cyanide. A small observational $i$:$n$ ratio for propyl cyanide may also be an indicator of the prevalence of such conditions.

The new models suggest, in contrast with the models of Belloche et al. (2014), that the main formation mechanism for $i$-C$_3$H$_7$CN is not the addition of CN to the C$_3$H$_7$ radical, but rather the addition of CH$_3$ to the secondary radical form of C$_2$H$_4$CN. This latter species is produced rapidly in the new network, through the CN + C$_2$H$_2$ and CN + C$_2$H$_4$ reactions, both of which are essentially barrierless and are the main reactions contributing to the production of vinyl and ethyl cyanide.

These new reactions involving CN make their strongest contribution to the overall production of nitriles here, at the level of C$_2$H$_3$CN and C$_2$H$_5$CN, rather than in the formation of propyl or butyl cyanide. However, if larger barriers to CN addition are assumed, with values chosen to agree with the barriers for equivalent reactions involving atomic H, an even larger ratio of {\em normal} to {\em iso} propyl cyanide is produced (although with smaller absolute abundances of each). However, this behavior is particularly associated with the larger barrier to the CN + C$_2$H$_2$ $\rightarrow$ C$_2$H$_2$CN reaction as compared with the reaction involving C$_2$H$_4$. This imbalance allows vinyl cyanide to be bypassed on the way to ethyl cyanide, bypassing also the secondary-radical form of C$_2$H$_4$CN that preferentially forms upon hydrogenation of C$_2$H$_3$CN by H. It is therefore possible that an even lower barrier for reaction 27 and a high barrier to reaction 21 could produce an even more extreme $n$:$i$ ratio in favor of $n$. The importance of these barriers to interstellar nitrile production might become clearer with experimental measurements down to lower temperatures, and/or values obtained for reactions specifically on ice surfaces.

In contrast to the propyl cyanide results, the models suggest that a branched form of butyl cyanide, $s$-C$_4$H$_9$CN, is always dominant over its straight-chain isomer. With a modeled abundance rising as high as $\sim$10$^{-8}$ with respect to H$_2$, it is also a prime candidate for detection toward Sgr B2(N). The ratio of the total abundance of all forms of butyl cyanide versus all forms of propyl cyanide, as predicted by the models, is on the order of unity, ranging from around 0.5--2.

Furthermore, as shown in Table 7, the ratio of maximum values of branched versus straight-chain forms of butyl cyanide reaches at least 3 in all models, while for the similar pentane molecules, this ratio is at least 2.6. It is quite plausible, therefore, to expect that molecules composed of at least 4 carbons plus one other functional group will in general show a bias toward branched structures, with the most abundant being the {\em sec} form. Based on the two data-points provided by the modeling of propyl and butyl cyanide, we might expect yet larger molecules to show an even stronger bias towards branching. The predicted peak total abundances of the two molecules are approximately equal within each model; however, the temperatures of peak abundance are expected to be higher for the larger species (due to larger grain-surface binding energies), meaning that observational searches could be limited by the size of the emission region. The larger partition functions (including rotational, vibrational, and often conformational components) associated with greater molecular complexity also result in relatively weaker emission from individual lines of such molecules (while the actual determination of such spectroscopic parameters for new molecules presents another, more immediate, challenge).

The abundance of OH in the ices could make alcohols good candidates for branched structures, assuming the temperature-dependent mobility of the OH radicals allows similar mechanisms to obtain as for CN. High-pressure gas-phase measurements of the OH+C$_2$H$_2$$\rightarrow$C$_2$H$_2$OH reaction indicate a barrier on the order of several hundred K (Mckee et al. 2007), which is likely to be low enough to make insertion of OH competitive on interstellar grain surfaces (due to competition with surface diffusion). In that case, this may be an efficient mechanism for the formation of vinyl alcohol on dust grains. It would also suggest that similar branched to straight-chain ratios might be obtained for the butyl alcohols as for the butyl cyanides, and an expected $i$:$n$ ratio for C$_3$H$_7$OH on the order of unity or a little less. 

\textit{Normal-} and \textit{iso-}propanol (C$_3$H$_7$OH) are not detected in the EMoCA survey toward Sgr~B2(N2). \citet{Mueller16} obtained column density upper limits of $2.6\times10^{17}$ cm$^{-2}$ and $9.3\times10^{16}$ cm$^{-2}$, respectively. These upper limits imply that \textit{n-}C$_3$H$_7$OH and \textit{i-}C$_3$H$_7$OH are at least 8 and 22 times less abundant than ethanol, respectively \citep[][]{Mueller16}. These non-detections do not allow us to draw conclusions as to which form of propanol prevails in Sgr~B2(N2). More sensitive observations will be needed to make further progress.

The low barriers to CN and OH insertion into multiple carbon bonds provide a plausible explanation for the production of {\em straight-chain} interstellar molecules, but other functional groups are unlikely to have such favorable barriers. Nevertheless, the addition of an H-atom to, for example, acetylene, is still a highly possible outcome for interstellar surface chemistry. In such a case, the likely product is a radical with the unpaired electron located at the secondary carbon, making the formation of a branched structure -- rather than a straight chain -- all the more probable. While they may be more abundant overall, nitriles and alcohols may indeed be the {\em least} branched molecules present in the ISM. 

Cosmic rays, and the secondary UV field they induce through collisions with H$_2$, are important to both the formation and destruction of complex organic molecules. Photodissociation of surface molecules leads to the production of radicals, which may react to form larger species.  Furthermore, cosmic-ray ionization is responsible for the production of molecular ions in the gas phase, which may protonate complex molecules, leading ultimately to their destruction. Unfortunately, the cosmic-ray ionization rate for Sgr B2(N2) is poorly constrained by observations. van der Tak et al. (2006) determined H$_3$O$^+$/H$_2$O ratios toward Sgr B2(OH), indicating a cosmic-ray ionization rate of $\zeta \simeq 4 \times 10^{-16}$ s$^{-1}$, with an uncertainty of a factor of 4; however, their value for the Sgr B2(M) core was an order of magnitude smaller. Assuming that the Sgr B2(M) value is an appropriate proxy for that of Sgr B2(N2), then the canonical interstellar value of $\zeta$ = $1.3 \times 10^{-17}$ s$^{-1}$ used in the models falls within the observational uncertainty, and is unlikely to be inaccurate by more than a factor of a few. Also, the large post-collapse visual extinction used in the models (as indeed confirmed by the observational H$_2$ column density of around $5 \times 10^{24}$ cm$^{-2}$) is likely to lead to some further attenuation of the CR field. Other sources of ionization affecting the chemistry in the hot core are unlikely; for example, there is no evidence for an embedded X-ray source in Sgr B2(N) \citep{Hong16}. While an embedded UCHII region does exist within Sgr B2(N2) \citep{DePree15}, with an angular size of $\sim$0.12'', the gas probed by our EMoCA observations has a scale size of 1''; a rudimentary calculation would indicate an intervening visual extinction on the order of at least several thousand magnitudes, far beyond any value at which the UV field associated with the UCHII region might conceivably affect the chemistry traced by our EMoCA observations.

The antagonistic nature of the interplay between CR-induced formation and destruction of complex organic molecules means that the influence of a modest inaccuracy in the cosmic-ray ionization rate should be small, for a given warm-up timescale. However, it is the overall fluence (i.e. time-integrated flux) of cosmic rays during the model run that will be important in both cases. One may therefore consider the three timescales tested in this study as an exploration of this parameter space, although the variations in timescale used here are likely larger than any inaccuracy in cosmic-ray ionization rate. A full investigation of the sensitivity of the hot-core models to these parameters is beyond the scope of this work, but is already under way.



\section{Conclusions}

\label{conclusions}

The simulations presented here make several strong predictions regarding the 
abundances of branched carbon-chain molecules in star-forming regions. 
The predictions specifically concerning butyl cyanide 
suggest that the EMoCA survey fell only a little short of detection of
this molecule. This prompted us to request new ALMA observations toward 
Sgr~B2(N), in order to search for this molecule; the observations are 
on-going. It also led us to seek to characterize the rotational spectra of
$s-$ and $i-$butyl cyanide, to allow for an astronomical search for these two 
branched isomers as well; the measurements and spectroscopic analyses are currently 
on-going at the University of Cologne.

Below, we list our main conclusions:

\begin{enumerate}
\item The observed {\em iso-} to {\em normal}-propyl cyanide ratio determined for Sgr B2(N2) by \citet{Belloche14} is well reproduced by the chemical models, with medium to fast warm-up timescales producing the best agreement. Absolute abundances for this molecule are also most accurately reproduced with the medium-timescale warm up.

\item Predicted peak fractional abundances of all forms of butyl cyanide are approximately equal to those obtained for propyl cyanide, within a factor of two. However, the distribution of any putative emission from butyl cyanide in Sgr B2(N2) is likely to be somewhat more compact than for propyl cyanide, due to the higher desorption temperature expected for this molecule.

\item The {\em sec} form of butyl cyanide is expected to be the most abundant, followed by {\em normal} and {\em iso}, which should have similar abundances. The {\em tert} form of butyl cyanide is significantly less abundant than the others, and is therefore likely to be a poor candidate for detection.

\item Reactions between the CN radical and hydrocarbons with multiple carbon-carbon bonds appear important to the chemistry of nitriles, while similar reactions involving OH may also be influential in the production of branched alcohols. Experimental or computational determinations of reaction rates/barriers down to low temperatures and/or on ice surfaces would help to constrain the models.

\item The effect of CN addition to hydrocarbons is to produce more of the straight-chain nitriles (with a similar effect likely for OH addition). Molecules for which similar formation mechanisms would involve much higher activation energy barriers are therefore likely to show greater branching, but with lower overall production rates. Nitriles and alcohols may therefore be some of the {\em least} branched molecules present in the ISM, as compared with species of similar size.

\item The production of the largest alkanes (butane and pentane) occurs mostly through the removal of an H-atom from a smaller alkane or the addition of H to an unsaturated hydrocarbon, followed by the addition of a methyl group. The addition of a methyl group (or sometimes an ethyl group) to a larger radical is often the most important mechanism for nitrile production, in those cases where the addition of CN to a hydrocarbon does not favor the formation of the required structure. It is likely that such mechanisms could play an important role in the formation of other large organic molecules on dust grains.

\end{enumerate}



\begin{acknowledgements}

The authors thank the referee for helpful comments.
We thank E. Willis for the creation of molecule images.
This work has been supported in part by the Deutsche 
Forschungsgemeinschaft through the collaborative research grant SFB 956 
``Conditions and Impact of Star Formation,'' project area B3.
This paper makes use of the following ALMA data:
ADS/JAO.ALMA\#2011.0.00017.S, ADS/JAO.ALMA\#2012.1.00012.S. 
ALMA is a partnership of ESO (representing its member states), NSF (USA) and 
NINS (Japan), together with NRC (Canada), NSC and ASIAA (Taiwan), and KASI 
(Republic of Korea), in cooperation with the Republic of Chile. The Joint ALMA 
Observatory is operated by ESO, AUI/NRAO and NAOJ.
The interferometric data are available in the ALMA archive at 
https://almascience.eso.org/aq/.

\end{acknowledgements}


\begin{table*}
\begin{center}
\caption{Physical quantities of new or related chemical species.}
\label{tab-species}
\renewcommand{\arraystretch}{1.0}
\small
\begin{tabular}[t]{lrrl}
\hline \hline
Species &  Binding energy & Enthalpy of formation, $\Delta H_{f}$(298 K) & Notes  \\
           &    (K)              &  (kcal mol$^{-1}$) \\
\hline

CH$_3$CN        &     4680      &       +17.70     &   \smallskip \\

C$_2$H$_2$CN      &      4187      &      +105.84     &    \\
C$_2$H$_3$CN      &      4637      &       +42.95     &    \\
$\overset{\centerdot}{\mathrm{C}}$H$_2$CH$_2$CN      &      5087      &       +55.13     &       Based on $\overset{\centerdot}{\mathrm{C}}$H$_2$CH$_2$CH$_3$ - CH$_3$$\overset{\centerdot}{\mathrm{C}}$HCH$_3$ \\
CH$_3$$\overset{\centerdot}{\mathrm{C}}$HCN      &      5087      &       +53.23     &         \\
C$_2$H$_5$CN      &      5537      &       +12.71     &        \smallskip  \\

$\overset{\centerdot}{\mathrm{C}}$H$_2$CH$_2$CH$_2$CN      &      6787      &       +56.38     &       Based on C$_3$H$_8$ - $\overset{\centerdot}{\mathrm{C}}$H$_2$CH$_2$CH$_3$ \\
CH$_3$$\overset{\centerdot}{\mathrm{C}}$HCH$_2$CN      &     6787      &       +54.48     &       Based on C$_3$H$_8$ - CH$_3$$\overset{\centerdot}{\mathrm{C}}$HCH$_3$  \\
CH$_3$CH$_2$$\overset{\centerdot}{\mathrm{C}}$HCN      &     6787      &       +54.48     &       Based on C$_3$H$_8$ - CH$_3$$\overset{\centerdot}{\mathrm{C}}$HCH$_3$ \\
$n$-C$_3$H$_7$CN      &      7237      &        +7.46     &  \smallskip   \\

$\overset{\centerdot}{\mathrm{C}}$H$_2$CH(CH$_3$)CN      &     6787      &       +54.36     &       Based on C$_3$H$_8$ - $\overset{\centerdot}{\mathrm{C}}$H$_2$CH$_2$CH$_3$  \\
CH$_3$$\overset{\centerdot}{\mathrm{C}}$(CH$_3$)CN      &     6787      &       +52.46     &       Based on C$_3$H$_8$ - CH$_3$$\overset{\centerdot}{\mathrm{C}}$HCH$_3$  \\

$i$-C$_3$H$_7$CN      &     7237      &        +5.44     &                         \smallskip       \\

CH$_3$CH$_2$$\overset{\centerdot}{\mathrm{C}}$HCH$_2$CN      &      8487      &      ---     &    \\
$n$-C$_4$H$_9$CN      &      8937      &        +2.65     &    \\
$i$-C$_4$H$_9$CN      &     8937      &        +0.58     &       $s$-C$_4$H$_9$CN  \\
$s$-C$_4$H$_9$CN      &     8937      &        +0.58     &    \\
$t$-C$_4$H$_9$CN      &     8937      &        --0.79     &   \smallskip  \\

$\overset{\centerdot}{\mathrm{C}}$H$_2$CH$_2$CH$_3$    &    5637      &      +23.90   &   \\
CH$_3$$\overset{\centerdot}{\mathrm{C}}$HCH$_3$    &    5637        &          +22.00           &   \\                       
C$_3$H$_8$    &     6087          &            --25.02    & \smallskip  \\

$\overset{\centerdot}{\mathrm{C}}$H$_2$CH$_2$CH$_2$CH$_3$    &      7337          &          +17.90      &      Based on $\overset{\centerdot}{\mathrm{C}}$H$_2$CH$_2$CH$_3$ - CH$_3$$\overset{\centerdot}{\mathrm{C}}$HCH$_3$H  \\
CH$_3$$\overset{\centerdot}{\mathrm{C}}$HCH$_2$CH$_3$     &     7337         &            +16.00       &       \\
$n$-C$_4$H$_{10}$      &    7787             &         -30.03    &   \smallskip  \\

$\overset{\centerdot}{\mathrm{C}}$H$_2$CH(CH$_3$)CH$_3$      &       7337      &       +17.00     &                                      \\
CH$_3$$\overset{\centerdot}{\mathrm{C}}$(CH$_3$)CH$_3$      &      7337      &       +15.10     &       Based on $\overset{\centerdot}{\mathrm{C}}$H$_2$CH$_2$CH$_3$ - CH$_3$$\overset{\centerdot}{\mathrm{C}}$HCH$_3$  \\
$i$-C$_4$H$_{10}$      &      7787      &       --32.07     &                         \smallskip       \\

$n$-C$_5$H$_{12}$      &       9487      &       --35.08     &    \\
$i$-C$_5$H$_{12}$      &      9487      &       --36.73     &    \\
{\em neo}-C$_5$H$_{12}$      &      9487      &       --40.14     &    \\
\hline
\end{tabular}
\end{center}
\tablefoot{
Dots indicate which carbon atom hosts the radical site (i.e. an unpaired electron), where appropriate. As in previous models, binding energies are representative of physisorption on an amorphous water ice surface.} Enthalpies of formation obtained from the NIST WebBook database; where not available, estimates were adopted as described in the Notes column.
\end{table*}


\begin{table*}
\begin{center}
\caption{Grain-surface/ice-mantle reactions involved in the formation of butane, pentane, and related radicals.}
\label{tab-reactions}
\renewcommand{\arraystretch}{1.0}
\small
\begin{tabular}[t]{rcrclcrclcrcl}
\hline \hline
\# && Reaction &&&&&&&& $E_{A}$ (K) && Ref. \\
\hline
1 && H   &  +  &  C$_2$H$_2$  &  $\rightarrow$  & $\overset{\centerdot}{\mathrm{C}}$HCH$_2$  & & & & 1300 && $a$  \\
2 && H   &  +  &  C$_2$H$_4$  &  $\rightarrow$  & $\overset{\centerdot}{\mathrm{C}}$H$_2$CH$_3$  & & & & 605 && $b$  \\
3 && H   &  +  &  C$_3$H$_6$  &  $\rightarrow$  & CH$_3$$\overset{\centerdot}{\mathrm{C}}$HCH$_3$  & & & & 619 && $c$  \\
4 && H   &  +  &  C$_3$H$_6$  &  $\rightarrow$  & $\overset{\centerdot}{\mathrm{C}}$H$_2$CH$_2$CH$_3$  & & & & 1320 && $c$  \\
5 &&  H  &  +  &  C$_4$H$_8$   &  $\rightarrow$  &  CH$_3$$\overset{\centerdot}{\mathrm{C}}$HCH$_2$CH$_3$ & & & & 619 && $d$ \\
6 &&  H  &  +  &  C$_4$H$_8$   &  $\rightarrow$  & $\overset{\centerdot}{\mathrm{C}}$H$_2$CH$_2$CH$_2$CH$_3$  & & & & 1320 && $d$ \\
\,\vspace{-2mm}\\

7 && $\overset{\centerdot}{\mathrm{C}}$H$_3$            &  +  & $\overset{\centerdot}{\mathrm{C}}$H$_2$CH$_2$CH$_3$  &  $\rightarrow$  &  $n$-C$_4$H$_{10}$ & & & & --  \\
8 && $\overset{\centerdot}{\mathrm{C}}$H$_2$CH$_3$  &  +  & $\overset{\centerdot}{\mathrm{C}}$H$_2$CH$_3$  &  $\rightarrow$  &  $n$-C$_4$H$_{10}$ & & & & --  \\
9 && H                                                                    &  +  & $\overset{\centerdot}{\mathrm{C}}$H$_2$CH$_2$CH$_2$CH$_3$  &  $\rightarrow$  &  $n$-C$_4$H$_{10}$ & & & & --  \\
10&& H                                                                    &  +  & CH$_3$$\overset{\centerdot}{\mathrm{C}}$HCH$_2$CH$_3$  &  $\rightarrow$  &  $n$-C$_4$H$_{10}$ & & & & --  \\
\,\vspace{-2mm}\\

11&& $\overset{\centerdot}{\mathrm{C}}$H$_3$            &  +  & CH$_3$$\overset{\centerdot}{\mathrm{C}}$HCH$_3$             &  $\rightarrow$  &  $i$-C$_4$H$_{10}$ & & & & --  \\
12&& H                                                                    &  +  & CH$_3$$\overset{\centerdot}{\mathrm{C}}$(CH$_3$)CH$_3$  &  $\rightarrow$  &  $i$-C$_4$H$_{10}$ & & & & --  \\
13&& \"{C}H$_2$                                                      &  +  &  CH$_3$$\overset{\centerdot}{\mathrm{C}}$HCH$_3$            &  $\rightarrow$  &  $\overset{\centerdot}{\mathrm{C}}$H$_2$CH(CH$_3$)CH$_3$ & & & & --  \\
14&& H                                                                    &  +  & $\overset{\centerdot}{\mathrm{C}}$H$_2$CH(CH$_3$)CH$_3$  &  $\rightarrow$  &  $i$-C$_4$H$_{10}$ & & & & --  \\
\,\vspace{-2mm}\\

15&& $\overset{\centerdot}{\mathrm{C}}$H$_3$            &  +  & $\overset{\centerdot}{\mathrm{C}}$H$_2$CH$_2$CH$_2$CH$_3$  &  $\rightarrow$  &  $n$-C$_5$H$_{12}$ & & & & --  \\
16&& $\overset{\centerdot}{\mathrm{C}}$H$_2$CH$_3$  &  +  & $\overset{\centerdot}{\mathrm{C}}$H$_2$CH$_2$CH$_3$  &  $\rightarrow$  &  $n$-C$_5$H$_{12}$ & & & & --  \\
\,\vspace{-2mm}\\

17&& $\overset{\centerdot}{\mathrm{C}}$H$_3$            &  +  & CH$_3$$\overset{\centerdot}{\mathrm{C}}$HCH$_2$CH$_3$  &  $\rightarrow$  &  $i$-C$_5$H$_{12}$ & & & & --  \\
18&& $\overset{\centerdot}{\mathrm{C}}$H$_2$CH$_3$  &  +  & CH$_3$$\overset{\centerdot}{\mathrm{C}}$HCH$_3$  &  $\rightarrow$  &  $i$-C$_5$H$_{12}$ & & & & --  \\
\,\vspace{-2mm}\\

19&& $\overset{\centerdot}{\mathrm{C}}$H$_3$            &  +  & CH$_3$$\overset{\centerdot}{\mathrm{C}}$(CH$_3$)CH$_3$  &  $\rightarrow$  &  {\em neo}-C$_5$H$_{12}$ & & & & --  \\

\hline
\end{tabular}
\end{center}
\tablefoot{
Dots indicate which carbon atom hosts the radical site (i.e. an unpaired electron), where appropriate. For brevity, we omit most hydrogen-addition steps to form ethane and propane. Reactions between simple radicals to form ethane, propane, or their precursors are similarly omitted from the table. Dashes indicate an assumed activation energy barrier of zero.
\tablefoottext{a}{Baulch et al. (1992)}
\tablefoottext{b}{\citet{Michael2005}}
\tablefoottext{c}{\citet{Curran2006}}
\tablefoottext{d}{Adopted value from equivalent C$_3$H$_6$ reaction}
}
\end{table*}


\begin{table*}

\begin{center}
\caption{Grain-surface/ice-mantle reactions involved in the formation of vinyl, ethyl, propyl, and butyl cyanide.}
\label{tab-reactions2}
\renewcommand{\arraystretch}{1.0}
\small
\begin{tabular}[t]{rcrclcrclcrcl}
\hline \hline
\# && Reaction &&&&&&&& $E_{A}$ (K) && Ref. \\
\hline
20&& H   &  +  &  HC$_3$N  &  $\rightarrow$  &  C$_2$H$_2$CN & & & & 1710 && $e$  \\
21&& $\overset{\centerdot}{\mathrm{C}}$N   &  +  & C$_2$H$_2$   &  $\rightarrow$  &  C$_2$H$_2$CN & & & & (i) 31, (ii) 1300 && $f$, $g$ \\
22&& H   &  +  &  C$_2$H$_2$CN  &  $\rightarrow$  &  C$_2$H$_3$CN & & & & --  \\
23&& $\overset{\centerdot}{\mathrm{C}}$N  &  +  &  $\overset{\centerdot}{\mathrm{C}}$HCH$_2$  &  $\rightarrow$  &  C$_2$H$_3$CN & & & & --  \\
\,\vspace{-2mm}\\

24&& H   &  +  &  C$_2$H$_3$CN  &  $\rightarrow$  & CH$_3$$\overset{\centerdot}{\mathrm{C}}$HCN  & & & & 619 && $h$ \\
25&& H   &  +  &  C$_2$H$_3$CN  &  $\rightarrow$  & $\overset{\centerdot}{\mathrm{C}}$H$_2$CH$_2$CN  & & & & 1320 && $h$ \\
26&& \"{C}H$_2$   &  +  &  $\overset{\centerdot}{\mathrm{C}}$H$_2$CN  &  $\rightarrow$  & $\overset{\centerdot}{\mathrm{C}}$H$_2$CH$_2$CN  & & & & -- \\
27&& $\overset{\centerdot}{\mathrm{C}}$N   &  +  & C$_2$H$_4$   &  $\rightarrow$  & $\overset{\centerdot}{\mathrm{C}}$H$_2$CH$_2$CN  & & & & (i) 31, (ii) 605 && $f$, $g$ \\
28&& H   &  +  &  CH$_3$$\overset{\centerdot}{\mathrm{C}}$HCN  &  $\rightarrow$  & C$_2$H$_5$CN  & & & & --  \\
29&& H   &  +  &  $\overset{\centerdot}{\mathrm{C}}$H$_2$CH$_2$CN  &  $\rightarrow$  & C$_2$H$_5$CN  & & & & --  \\
30&& $\overset{\centerdot}{\mathrm{C}}$H$_3$  &  +  &  $\overset{\centerdot}{\mathrm{C}}$H$_2$CN  &  $\rightarrow$  & C$_2$H$_5$CN  & & & & --  \\
31&& $\overset{\centerdot}{\mathrm{C}}$N   &  +  & C$_2$H$_6$   &  $\rightarrow$  & $\overset{\centerdot}{\mathrm{C}}$H$_2$CH$_3$ & + & HCN & & 0 && $f$ \\
32&& $\overset{\centerdot}{\mathrm{C}}$N  &  +  &  $\overset{\centerdot}{\mathrm{C}}$H$_2$CH$_3$  &  $\rightarrow$  & C$_2$H$_5$CN  & & & & --  \\
\,\vspace{-2mm}\\

33&& $\overset{\centerdot}{\mathrm{C}}$H$_3$                      &  +  & $\overset{\centerdot}{\mathrm{C}}$H$_2$CH$_2$CN   &  $\rightarrow$  &  $n$-C$_3$H$_7$CN & & & & --  \\
34&& $\overset{\centerdot}{\mathrm{C}}$H$_2$CH$_3$            &  +  & $\overset{\centerdot}{\mathrm{C}}$H$_2$CN             &  $\rightarrow$  &  $n$-C$_3$H$_7$CN & & & & --  \\
35&& $\overset{\centerdot}{\mathrm{C}}$H$_2$CH$_2$CH$_3$  &  +  & $\overset{\centerdot}{\mathrm{C}}$N                       &  $\rightarrow$  &  $n$-C$_3$H$_7$CN & & & & --  \\
36&& \"{C}H$_2$        &  +  & $\overset{\centerdot}{\mathrm{C}}$H$_2$CH$_2$CN    &  $\rightarrow$  &  $\overset{\centerdot}{\mathrm{C}}$H$_2$CH$_2$CH$_2$CN & & & & --  \\
37&& H                     &  +  & $\overset{\centerdot}{\mathrm{C}}$H$_2$CH$_2$CH$_2$CN   &  $\rightarrow$  &  $n$-C$_3$H$_7$CN & & & & --  \\
38&&  $\overset{\centerdot}{\mathrm{C}}$N  &  +  &  C$_3$H$_6$   &  $\rightarrow$  &  CH$_3$$\overset{\centerdot}{\mathrm{C}}$HCH$_2$CN & & & & (i) 0, (ii) 619 && $i$, $h$  \\
39&& H                     &  +  &  CH$_3$$\overset{\centerdot}{\mathrm{C}}$HCH$_2$CN  &  $\rightarrow$  &  $n$-C$_3$H$_7$CN & & & & --  \\
40&& H                     &  +  &  CH$_3$CH$_2$$\overset{\centerdot}{\mathrm{C}}$HCN  &  $\rightarrow$  &  $n$-C$_3$H$_7$CN & & & & --  \\
\,\vspace{-2mm}\\

41&& $\overset{\centerdot}{\mathrm{C}}$H$_3$                       &  +  & CH$_3$$\overset{\centerdot}{\mathrm{C}}$HCN   &  $\rightarrow$  &  $i$-C$_3$H$_7$CN & & & & --  \\
42&& CH$_3$$\overset{\centerdot}{\mathrm{C}}$HCH$_3$         &  +  & $\overset{\centerdot}{\mathrm{C}}$N                  &  $\rightarrow$  &  $i$-C$_3$H$_7$CN & & & & --  \\
43&& \"{C}H$_2$     &  +  &  CH$_3$$\overset{\centerdot}{\mathrm{C}}$HCN               &  $\rightarrow$  &  $\overset{\centerdot}{\mathrm{C}}$H$_2$CH(CH$_3$)CN & & & & --  \\
44&& H                  &  +  &  $\overset{\centerdot}{\mathrm{C}}$H$_2$CH(CH$_3$)CN   &  $\rightarrow$  &  $i$-C$_3$H$_7$CN & & & & --  \\
45&& H                  &  +  & CH$_3$$\overset{\centerdot}{\mathrm{C}}$(CH$_3$)CN    &  $\rightarrow$  &  $i$-C$_3$H$_7$CN & & & & --  \\
\,\vspace{-2mm}\\

46&& $\overset{\centerdot}{\mathrm{C}}$H$_3$            &  +  & $\overset{\centerdot}{\mathrm{C}}$H$_2$CH$_2$CH$_2$CN   &  $\rightarrow$  &  $n$-C$_4$H$_9$CN & & & & --  \\
47&& $\overset{\centerdot}{\mathrm{C}}$H$_2$CH$_3$  &  +  & $\overset{\centerdot}{\mathrm{C}}$H$_2$CH$_2$CN             &  $\rightarrow$  &  $n$-C$_4$H$_9$CN & & & & --  \\
48&& $\overset{\centerdot}{\mathrm{C}}$H$_2$CH$_2$CH$_3$  &  +  & $\overset{\centerdot}{\mathrm{C}}$H$_2$CN             &  $\rightarrow$  &  $n$-C$_4$H$_9$CN & & & & --  \\
49&& $\overset{\centerdot}{\mathrm{C}}$H$_2$CH$_2$CH$_2$CH$_3$  &  +  & $\overset{\centerdot}{\mathrm{C}}$N             &  $\rightarrow$  &  $n$-C$_4$H$_9$CN & & & & --  \\
50&&  $\overset{\centerdot}{\mathrm{C}}$N  &  +  &  C$_4$H$_8$   &  $\rightarrow$  &  CH$_3$CH$_2$$\overset{\centerdot}{\mathrm{C}}$HCH$_2$CN & & & & (i) 0, (ii) 619 && $i$, $h$ \\
51&& H                     &  +  &  CH$_3$CH$_2$$\overset{\centerdot}{\mathrm{C}}$HCH$_2$CN  &  $\rightarrow$  &  $n$-C$_4$H$_9$CN & & & & --  \\
\,\vspace{-2mm}\\

52&& $\overset{\centerdot}{\mathrm{C}}$H$_3$            &  +  & CH$_3$$\overset{\centerdot}{\mathrm{C}}$HCH$_2$CN   &  $\rightarrow$  &  $i$-C$_4$H$_9$CN & & & & --  \\
53&& CH$_3$$\overset{\centerdot}{\mathrm{C}}$HCH$_3$  &  +  & $\overset{\centerdot}{\mathrm{C}}$H$_2$CN             &  $\rightarrow$  &  $i$-C$_4$H$_9$CN & & & & --  \\
54&& $\overset{\centerdot}{\mathrm{C}}$H$_2$CH(CH$_3$)CH$_3$  &  +  & $\overset{\centerdot}{\mathrm{C}}$N           &  $\rightarrow$  &  $i$-C$_4$H$_9$CN & & & & --  \\
\,\vspace{-2mm}\\

55&& $\overset{\centerdot}{\mathrm{C}}$H$_3$            &  +  & CH$_3$CH$_2$$\overset{\centerdot}{\mathrm{C}}$HCN    &  $\rightarrow$  &  $s$-C$_4$H$_9$CN & & & & --  \\
56&& $\overset{\centerdot}{\mathrm{C}}$H$_3$            &  +  & $\overset{\centerdot}{\mathrm{C}}$H$_2$CH(CH$_3$)CN  &  $\rightarrow$  &  $s$-C$_4$H$_9$CN & & & & --  \\
57&& $\overset{\centerdot}{\mathrm{C}}$H$_2$CH$_3$  &  +  & CH$_3$$\overset{\centerdot}{\mathrm{C}}$HCN             &  $\rightarrow$  &  $s$-C$_4$H$_9$CN & & & & --  \\
58&& CH$_3$$\overset{\centerdot}{\mathrm{C}}$HCH$_2$CH$_3$  &  +  & $\overset{\centerdot}{\mathrm{C}}$N             &  $\rightarrow$  &  $s$-C$_4$H$_9$CN & & & & --  \\
\,\vspace{-2mm}\\

59&& $\overset{\centerdot}{\mathrm{C}}$H$_3$            &  +  & CH$_3$$\overset{\centerdot}{\mathrm{C}}$(CH$_3$)CN    &  $\rightarrow$  &  $t$-C$_4$H$_9$CN & & & & --  \\
60&& CH$_3$$\overset{\centerdot}{\mathrm{C}}$(CH$_3$)CH$_3$  &  +  & $\overset{\centerdot}{\mathrm{C}}$N             &  $\rightarrow$  &  $t$-C$_4$H$_9$CN & & & & --  \\

\hline
\end{tabular}
\end{center}
\tablefoot{
Dots indicate which carbon atom hosts the radical site (i.e. an unpaired electron), where appropriate; where the radical site is not indicated, no distinction is made in the model between structural isomers. Dashes indicate an assumed activation energy barrier of zero. Reactions 21, 27, 38, \& 50 each have two values shown, corresponding to the (i) standard and (ii) high-barrier models, respectively.
\tablefoottext{e}{Parker et al. (2004)}
\tablefoottext{f}{Based on gas-phase value}
\tablefoottext{g}{Adopted value from equivalent H-addition reaction}
\tablefoottext{h}{Adopted value from equivalent C$_3$H$_6$ reaction}
\tablefoottext{i}{Based on fit to \citet{Gannon2007} gas-phase data}
}
\end{table*}



\begin{table*}
\begin{center}
\caption{Grain-surface/ice-mantle hydrogen-abstraction reactions involved in the formation of butane, pentane, propyl cyanide, butyl cyanide, and related radicals.}
\label{tab-reactions3}
\renewcommand{\arraystretch}{1.0}
\small
\begin{tabular}[t]{rcrclcrclcrcl}
\hline \hline
\# && Reaction &&&&&&&& $E_{A}$ (K) && Ref. \\
\hline
61  && $\overset{\centerdot}{\mathrm{C}}$H$_2$OH &  +  &  CH$_3$CN            &   $\rightarrow$  & CH$_3$OH  &  +  & $\overset{\centerdot}{\mathrm{C}}$H$_2$CN                           &&        6200  && Evans-Polanyi  \\
62  && $\overset{\centerdot}{\mathrm{C}}$H$_2$OH &  +  &  C$_2$H$_5$CN      &  $\rightarrow$  &  CH$_3$OH  &  +  & $\overset{\centerdot}{\mathrm{C}}$H$_2$CH$_2$CN                 &&       6490  && Evans-Polanyi   \\
63  && $\overset{\centerdot}{\mathrm{C}}$H$_2$OH &  +  &  C$_2$H$_5$CN      &  $\rightarrow$  &  CH$_3$OH  &  +  & CH$_3$$\overset{\centerdot}{\mathrm{C}}$HCN                        &&       5990  && Evans-Polanyi   \\

64  && CH$_3$$\overset{\centerdot}{\mathrm{O}}$  &  +  &  CH$_3$CN             &  $\rightarrow$  &  CH$_3$OH  &  +  &       $\overset{\centerdot}{\mathrm{C}}$H$_2$CN                      &&       2070  && Evans-Polanyi   \\
65  && CH$_3$$\overset{\centerdot}{\mathrm{O}}$  &  +  &  C$_2$H$_5$CN       &  $\rightarrow$  &  CH$_3$OH &  +  &      $\overset{\centerdot}{\mathrm{C}}$H$_2$CH$_2$CN             &&       2340  && Evans-Polanyi   \\
66  && CH$_3$$\overset{\centerdot}{\mathrm{O}}$  &  +  &  C$_2$H$_5$CN       &  $\rightarrow$  &  CH$_3$OH &  +  &    CH$_3$$\overset{\centerdot}{\mathrm{C}}$HCN                      &&       1950  && Evans-Polanyi   \\
67  && CH$_3$$\overset{\centerdot}{\mathrm{O}}$  &  +  &  C$_3$H$_7$CN       &  $\rightarrow$  &  CH$_3$OH &  +  &    $\overset{\centerdot}{\mathrm{C}}$H$_2$CH$_2$CH$_2$CN     &&       3660  && Evans-Polanyi   \\
68  && CH$_3$$\overset{\centerdot}{\mathrm{O}}$  &  +  &  C$_3$H$_7$CN       &  $\rightarrow$  &  CH$_3$OH &  +  &  CH$_3$$\overset{\centerdot}{\mathrm{C}}$HCH$_2$CN              &&      3260  && Evans-Polanyi   \\
69  && CH$_3$$\overset{\centerdot}{\mathrm{O}}$  &  +  &  C$_3$H$_7$CN       &  $\rightarrow$  &  CH$_3$OH &  +  &  CH$_3$CH$_2$$\overset{\centerdot}{\mathrm{C}}$HCN              &&      3260  && Evans-Polanyi   \\
70  && CH$_3$$\overset{\centerdot}{\mathrm{O}}$  &  +  &  $i$-C$_3$H$_7$CN  &  $\rightarrow$  & CH$_3$OH  &  +  & CH$_3$$\overset{\centerdot}{\mathrm{C}}$(CH$_3$)CN               &&      3260  && Evans-Polanyi   \\
71  && CH$_3$$\overset{\centerdot}{\mathrm{O}}$  &  +  &  $i$-C$_3$H$_7$CN  &  $\rightarrow$  & CH$_3$OH  &  +  & $\overset{\centerdot}{\mathrm{C}}$H$_2$CH(CH$_3$)CN             &&      3660  && Evans-Polanyi   \\
72  && CH$_3$$\overset{\centerdot}{\mathrm{O}}$  &  +  &  C$_2$H$_6$          &  $\rightarrow$  &  CH$_3$OH  &  +  & $\overset{\centerdot}{\mathrm{C}}$H$_2$CH$_3$                       &&      3560  && Evans-Polanyi   \\
73  && CH$_3$$\overset{\centerdot}{\mathrm{O}}$  &  +  &  C$_3$H$_8$          &  $\rightarrow$  &  CH$_3$OH  &  +  & CH$_3$$\overset{\centerdot}{\mathrm{C}}$HCH$_3$                   &&      3260  && Evans-Polanyi   \\
74  && CH$_3$$\overset{\centerdot}{\mathrm{O}}$  &  +  &  C$_3$H$_8$          &  $\rightarrow$  &  CH$_3$OH  &  +  & $\overset{\centerdot}{\mathrm{C}}$H$_2$CH$_2$CH$_3$             &&     3660  && Evans-Polanyi   \\
75  && CH$_3$$\overset{\centerdot}{\mathrm{O}}$  &  +  &  C$_4$H$_{10}$      &  $\rightarrow$  &   CH$_3$OH &  +  &  $\overset{\centerdot}{\mathrm{C}}$H$_2$CH$_2$CH$_2$CH$_3$ &&      3450  && Evans-Polanyi   \\
76  && CH$_3$$\overset{\centerdot}{\mathrm{O}}$  &  +  &  C$_4$H$_{10}$      &  $\rightarrow$  &   CH$_3$OH &  +  &  CH$_3$$\overset{\centerdot}{\mathrm{C}}$HCH$_2$CH$_3$        &&     3060  && Evans-Polanyi   \\
77  && CH$_3$$\overset{\centerdot}{\mathrm{O}}$  &  +  &  $i$-C$_4$H$_{10}$  &  $\rightarrow$  &  CH$_3$OH  &  +  & CH$_3$$\overset{\centerdot}{\mathrm{C}}$(CH$_3$)CH$_3$        &&     3310  && Evans-Polanyi   \\
78  && CH$_3$$\overset{\centerdot}{\mathrm{O}}$  &  +  &  $i$-C$_4$H$_{10}$  &  $\rightarrow$  &  CH$_3$OH  &  +  & $\overset{\centerdot}{\mathrm{C}}$H$_2$CH(CH$_3$)CH$_3$      &&     3700  && Evans-Polanyi   \\

79  && $\overset{\centerdot}{\mathrm{N}}$H$_2$  &  +  &   CH$_3$CN              &  $\rightarrow$  &  NH$_3$   &  +  &  $\overset{\centerdot}{\mathrm{C}}$H$_2$CN                               &&     2680  && Evans-Polanyi   \\
80  && $\overset{\centerdot}{\mathrm{N}}$H$_2$  &  +  &   C$_2$H$_5$CN        &  $\rightarrow$  &  NH$_3$  &  +  &   CH$_3$$\overset{\centerdot}{\mathrm{C}}$HCN                           &&     2480  && Evans-Polanyi   \\
81  && $\overset{\centerdot}{\mathrm{N}}$H$_2$  &  +  &   C$_2$H$_5$CN        &  $\rightarrow$  &  NH$_3$  &  +  &   $\overset{\centerdot}{\mathrm{C}}$H$_2$CH$_2$CN                     &&     3280  && Evans-Polanyi   \\
82  && $\overset{\centerdot}{\mathrm{N}}$H$_2$  &  +  &   C$_3$H$_7$CN        &  $\rightarrow$  &  NH$_3$  &  +  &   CH$_3$$\overset{\centerdot}{\mathrm{C}}$HCH$_2$CN                 &&     5160  && Evans-Polanyi   \\
83  && $\overset{\centerdot}{\mathrm{N}}$H$_2$  &  +  &   C$_3$H$_7$CN        &  $\rightarrow$  &  NH$_3$  &  +  &   CH$_3$CH$_2$$\overset{\centerdot}{\mathrm{C}}$HCN                 &&     5160  && Evans-Polanyi   \\
84  && $\overset{\centerdot}{\mathrm{N}}$H$_2$  &  +  &   C$_3$H$_7$CN        &  $\rightarrow$  &  NH$_3$  &  +  &   $\overset{\centerdot}{\mathrm{C}}$H$_2$CH$_2$CH$_2$CN           &&     5980  && Evans-Polanyi   \\
85  && $\overset{\centerdot}{\mathrm{N}}$H$_2$  &  +  &   $i$-C$_3$H$_7$CN   &  $\rightarrow$  &  NH$_3$  &  +  &   $\overset{\centerdot}{\mathrm{C}}$H$_2$CH(CH$_3$)CN               &&     5980  && Evans-Polanyi   \\
86  && $\overset{\centerdot}{\mathrm{N}}$H$_2$  &  +  &   $i$-C$_3$H$_7$CN   &  $\rightarrow$  &  NH$_3$  &  +  &   CH$_3$$\overset{\centerdot}{\mathrm{C}}$(CH$_3$)CN                 &&     5160  && Evans-Polanyi   \\
87  && $\overset{\centerdot}{\mathrm{N}}$H$_2$  &  +  &   C$_2$H$_6$           &  $\rightarrow$  &  NH$_3$    &  +  &   $\overset{\centerdot}{\mathrm{C}}$H$_2$CH$_3$                        &&    5770  && Hennig \& Wagner (1995) \\
88  && $\overset{\centerdot}{\mathrm{N}}$H$_2$  &  +  &   C$_3$H$_8$           &  $\rightarrow$  &  NH$_3$    &  +  & CH$_3$$\overset{\centerdot}{\mathrm{C}}$HCH$_3$                      &&    5150  && Evans-Polanyi   \\
89  && $\overset{\centerdot}{\mathrm{N}}$H$_2$  &  +  &   C$_3$H$_8$           &  $\rightarrow$  &  NH$_3$    &  +  &  $\overset{\centerdot}{\mathrm{C}}$H$_2$CH$_2$CH$_3$               &&   5980  && Evans-Polanyi   \\
90  && $\overset{\centerdot}{\mathrm{N}}$H$_2$  &  +  &   C$_4$H$_{10}$       &  $\rightarrow$  &   NH$_3$    &  +  & $\overset{\centerdot}{\mathrm{C}}$H$_2$CH$_2$CH$_2$CH$_3$    &&    5550  && Evans-Polanyi   \\
91  && $\overset{\centerdot}{\mathrm{N}}$H$_2$  &  +  &   C$_4$H$_{10}$       &  $\rightarrow$  &   NH$_3$    &  +  & CH$_3$$\overset{\centerdot}{\mathrm{C}}$HCH$_2$CH$_3$           &&   4730  && Evans-Polanyi   \\
92  && $\overset{\centerdot}{\mathrm{N}}$H$_2$  &  +  &   $i$-C$_4$H$_{10}$  &  $\rightarrow$  &  NH$_3$    &  +  &  $\overset{\centerdot}{\mathrm{C}}$H$_2$CH(CH$_3$)CH$_3$         &&   6100  && Evans-Polanyi   \\
93  && $\overset{\centerdot}{\mathrm{N}}$H$_2$  &  +  &   $i$-C$_4$H$_{10}$  &  $\rightarrow$  &  NH$_3$    &  +  & CH$_3$$\overset{\centerdot}{\mathrm{C}}$(CH$_3$)CH$_3$            &&   5290  && Evans-Polanyi   \\

94  && $\overset{\centerdot}{\mathrm{O}}$H   &  +  &   CH$_3$CN                   &  $\rightarrow$  &  H$_2$O   &  +  &  $\overset{\centerdot}{\mathrm{C}}$H$_2$CN                                &&  500  && Kurylo \& Knable (1984) \\
95  && $\overset{\centerdot}{\mathrm{O}}$H   &  +  &   C$_2$H$_3$CN             &  $\rightarrow$  &  H$_2$O   &  +  &  C$_2$H$_2$CN                                                                         &&   4000  && Evans-Polanyi   \\
96  && $\overset{\centerdot}{\mathrm{O}}$H   &  +  &   C$_2$H$_5$CN             &  $\rightarrow$  &  H$_2$O   &  +  &  $\overset{\centerdot}{\mathrm{C}}$H$_2$CH$_2$CN                     &&   1200  && Estimate, based on \#103 \\
97  && $\overset{\centerdot}{\mathrm{O}}$H   &  +  &   C$_2$H$_5$CN             &  $\rightarrow$  &  H$_2$O   &  +  &  CH$_3$$\overset{\centerdot}{\mathrm{C}}$HCN                            &&  1000  && Estimate, based on \#104 \\
98  && $\overset{\centerdot}{\mathrm{O}}$H   &  +  &   C$_3$H$_7$CN             &  $\rightarrow$  &  H$_2$O   &  +  &  $\overset{\centerdot}{\mathrm{C}}$H$_2$CH$_2$CH$_2$CN           &&   1000  && Estimate, based on \#103 \\
99  && $\overset{\centerdot}{\mathrm{O}}$H   &  +  &   C$_3$H$_7$CN             &  $\rightarrow$  &  H$_2$O   &  +  &  CH$_3$$\overset{\centerdot}{\mathrm{C}}$HCH$_2$CN                  &&  800  && Estimate, based on \#104 \\
100 && $\overset{\centerdot}{\mathrm{O}}$H   &  +  &   C$_3$H$_7$CN             &  $\rightarrow$  &  H$_2$O   &  +  &  CH$_3$CH$_2$$\overset{\centerdot}{\mathrm{C}}$HCN                  &&  800  && Estimate, based on \#104 \\
101  && $\overset{\centerdot}{\mathrm{O}}$H   &  +  &   $i$-C$_3$H$_7$CN        &  $\rightarrow$  &  H$_2$O   &  +  &  $\overset{\centerdot}{\mathrm{C}}$H$_2$CH(CH$_3$)CN                &&  1000  && Estimate, based on \#103 \\
102  && $\overset{\centerdot}{\mathrm{O}}$H   &  +  &   $i$-C$_3$H$_7$CN        &  $\rightarrow$  &  H$_2$O   &  +  &  CH$_3$$\overset{\centerdot}{\mathrm{C}}$(CH$_3$)CN                  &&  800  && Estimate, based on \#104 \\
103  && $\overset{\centerdot}{\mathrm{O}}$H   &  +  &   C$_2$H$_6$                &  $\rightarrow$  &  H$_2$O   &  +  &  $\overset{\centerdot}{\mathrm{C}}$H$_2$CH$_3$                          &&  1000  && Atkinson et al. 2001 \\
104  && $\overset{\centerdot}{\mathrm{O}}$H   &  +  &   C$_3$H$_8$                &  $\rightarrow$  &  H$_2$O   &  +  &  $\overset{\centerdot}{\mathrm{C}}$H$_2$CH$_2$CH$_3$                &&  1310  && Hu et al. (1997) \\
105  && $\overset{\centerdot}{\mathrm{O}}$H   &  +  &   C$_3$H$_8$                &  $\rightarrow$  &  H$_2$O   &  +  &  CH$_3$$\overset{\centerdot}{\mathrm{C}}$HCH$_3$                      &&   1120  && Hu et al. (1997) \\
106  && $\overset{\centerdot}{\mathrm{O}}$H   &  +  &   C$_4$H$_{10}$            &  $\rightarrow$  &   H$_2$O  &  +  &   $\overset{\centerdot}{\mathrm{C}}$H$_2$CH$_2$CH$_2$CH$_3$    &&   823  && Greiner (1970) \\
107  && $\overset{\centerdot}{\mathrm{O}}$H   &  +  &   C$_4$H$_{10}$           &  $\rightarrow$  &   H$_2$O  &  +  &   CH$_3$$\overset{\centerdot}{\mathrm{C}}$HCH$_2$CH$_3$            &&  428  && Greiner (1970) \\
108  && $\overset{\centerdot}{\mathrm{O}}$H   &  +  &   $i$-C$_4$H$_{10}$      &  $\rightarrow$  &  H$_2$O   &  +  &  $\overset{\centerdot}{\mathrm{C}}$H$_2$CH(CH$_3$)CH$_3$           &&  823  && Estimate, based on \#105 \\
109  && $\overset{\centerdot}{\mathrm{O}}$H   &  +  &   $i$-C$_4$H$_{10}$      &  $\rightarrow$  &  H$_2$O   &  +  &  CH$_3$$\overset{\centerdot}{\mathrm{C}}$(CH$_3$)CH$_3$             &&  428  && Estimate, based on \#106 \\

\hline
\end{tabular}
\end{center}
\tablefoot{
Dots indicate which carbon atom hosts the radical site (i.e. an unpaired electron), where appropriate. The origins of each adopted activation energy are indicated
}
\end{table*}


\begin{table*}
\begin{center}
\caption{Selection of protostellar ice compositions from the literature, and post-collapse model values, as a percentage of H$_2$O.}
\label{tab-ratios}
\renewcommand{\arraystretch}{1.0}
\small
\begin{tabular}[t]{lcccccc}
\hline \hline
Species & W33A $^a$ & NGC 7538 IRS 9 $^b$ & Sgr A* $^b$ & Typical & Typical & Collapse-phase \\
&&&& low-mass $^c$ & high-mass $^c$ & model values \\
\hline
CO                  &  8    &  16 &  $<$12 &  29    &   13                &  53      \\
CO$_2$           &  13   &  22 &  14       &  29   &   13                &  14    \\
CH$_4$           &  1.5  &  2   &  2        &  5   &    2                   &  2.9          \\
H$_2$CO         &  6    &  4   &  $<$3   &  $\leq$2   &   $\leq$2    &  1.8       \\
CH$_3$OH       &  18  &  5   &  $<$4   &  3   &   4                    &  6.9      \\
NH$_3$           &  15  &  13  &  20--30 &  5   &   5                    &  20       \\
\hline
\end{tabular}
\end{center}

\tablefoot{
\tablefoottext{a}{Gibb et al. (2000)}
\tablefoottext{b}{See Gibb et al. (2000) for original references}
\tablefoottext{c}{{\"O}berg et al. (2011)}
}
\end{table*}



\begin{table*}
\begin{center}
\caption{Peak gas-phase fractional abundances (with respect to H$_2$) and corresponding model temperatures for a selection of molecules.}
\label{tab-abuns}
\renewcommand{\arraystretch}{1.0}
\small
\begin{tabular}[t]{lrrrrrrrrrrrrrrrrr}

\hline \hline
 & \multicolumn{5}{c}{Fast} && \multicolumn{5}{c}{Medium} && \multicolumn{5}{c}{Slow} \\
\cline{2-6} \cline{8-12} \cline{14-18} \\

 & \multicolumn{2}{c}{Low E$_A$} && \multicolumn{2}{c}{High E$_A$} && \multicolumn{2}{c}{Low E$_A$} && \multicolumn{2}{c}{High E$_A$} && \multicolumn{2}{c}{Low E$_A$} && \multicolumn{2}{c}{High E$_A$} \\
\cline{2-3} \cline{5-6} \cline{8-9} \cline{11-12} \cline{14-15} \cline{17-18}  \\


Species & $X[i]$ & $T$  && $X[i]$ & $T$  && $X[i]$ & $T$ && $X[i]$ & $T$ && $X[i]$ & $T$ && $X[i]$ & $T$ \\
           &         & (K)   &&          & (K)   &&          & (K)   &&          & (K)   &&          & (K)   &&          & (K) \\

\hline

HCN                                   & 5.2(-7)  & 400 && 5.2(-7)  & 400 && 1.2(-6)   & 398 && 1.2(-6)    & 398 && 3.9(-6)  & 400 && 3.9(-6) & 400  \\
CH$_3$CN                           & 5.8(-9)  & 400 && 5.8(-9)  & 400 && 7.4(-9)   & 398 && 6.8(-9)    & 398 && 3.4(-9)  & 400 && 3.3(-9) & 400  \\
C$_2$H$_3$CN                    & 2.5(-9)  & 400 && 1.6(-9)   & 400 && 1.7(-8)  & 285 && 1.3(-8)    & 285 && 7.4(-9)   & 167 && 4.7(-9) & 171  \\
C$_2$H$_5$CN                    & 2.3(-8)  & 131 && 1.2(-8)   & 131 && 1.1(-7)  & 130 && 8.1(-8)    & 130 && 3.5(-8)   & 129 && 1.8(-8) & 129  \\
$n$-C$_3$H$_7$CN              & 5.6(-10) & 166 && 1.7(-9)   & 166 && 1.4(-8)  & 160 && 1.4(-8)    & 160 && 1.9(-9)   & 156 && 1.2(-9) & 156  \\
$i$-C$_3$H$_7$CN               & 1.7(-9)   & 164 && 7.6(-10) & 164 && 3.3(-9)   & 160 && 2.3(-9)    & 160 && 3.4(-9)  & 154 && 1.7(-9) & 154  \\
$n$-C$_4$H$_9$CN              & 2.9(-10) & 204 && 1.9(-10) & 204 && 4.6(-9)   & 196 && 4.4(-9)    & 196 && 2.0(-9)  & 190 && 1.2(-9) & 190  \\
$i$-C$_4$H$_9$CN               & 2.4(-10) & 204 && 3.0(-10) & 202 && 2.8(-9)   & 196 && 2.6(-9)    & 196 && 3.8(-9)  & 190 && 2.7(-9) & 190  \\
$s$-C$_4$H$_9$CN              & 1.3(-9)   & 204 && 7.9(-10) & 204 && 1.1(-8)   & 196 && 1.1(-8)   & 196 && 3.7(-9)   & 190 && 2.1(-9) & 190  \\
$t$-C$_4$H$_9$CN              & 1.2(-11)  & 204 && 5.2(-12) & 204 && 7.1(-11) & 196 && 5.8(-11)  & 196 && 2.0(-10) & 190 && 1.0(-10) & 190  \\
C$_3$H$_6$                       & 9.7(-9)   & 400 && 9.5(-9)   & 400 && 7.3(-8)   & 398 && 7.3(-8)   & 398 && 1.7(-7)   & 395 && 1.7(-7) & 395  \\
C$_3$H$_8$                       & 2.8(-7)   & 143 && 2.8(-7)   & 145 && 6.3(-7)   & 143 && 6.4(-7)   & 143 && 1.8(-6)   & 139 && 1.7(-6) & 139  \\
$n$-C$_4$H$_1$$_0$           & 2.3(-8)   & 182 && 2.6(-8)   & 182 && 1.7(-7)   & 182 && 1.7(-7)   & 183 && 2.5(-7)   & 177 && 2.5(-7) & 177  \\
$i$-C$_4$H$_1$$_0$            & 3.0(-9)   & 182 && 3.2(-9)   & 184 && 9.2(-9)   & 185 && 9.3(-9)   & 185 && 9.1(-8)   & 177 && 8.9(-8) & 177  \\
$n$-C$_5$H$_1$$_2$           & 1.2(-10) & 226 && 1.4(-10)  & 226 && 1.3(-9)   & 220 && 1.4(-9)   & 220 && 1.2(-8)   & 212 && 1.2(-8) & 212  \\
$i$-C$_5$H$_1$$_2$            & 4.0(-9)   & 224 && 4.4(-9)   & 224 && 4.0(-9)   & 218 && 4.1(-9)   & 218 && 2.5(-8)   & 210 && 2.5(-8) & 210  \\
{\em neo}-C$_5$H$_1$$_2$  & 3.4(-11) & 224 && 3.7(-11)  & 228 && 1.4(-10) & 222 && 1.4(-10) & 222 && 5.9(-9)   & 214 && 5.9(-9) & 214  \\
H$_2$O                             & 3.3(-4)   & 396 && 3.3(-4)   & 396 && 3.4(-4)   & 398 && 3.4(-4)   & 398 && 3.6(-4)   & 400 && 3.5(-4) & 400  \\
CH$_3$OH                         & 2.4(-5)   & 131 && 2.4(-5)   & 131 && 2.1(-5)   & 130 && 2.1(-5)   & 130 && 1.1(-5)   & 129 && 1.1(-5) & 129  \\
CH$_3$OCH$_3$                 & 1.2(-7)   & 400 && 1.2(-7)   & 400 && 2.0(-7)   & 362 && 2.0(-7)   & 362 && 1.3(-7)   & 161 && 1.3(-7) & 161  \\
HCOOCH$_3$                     & 3.1(-7)   & 130 && 3.1(-7)   & 130 && 7.8(-8)   &  125 && 7.8(-8)  & 125 && 4.1(-10)  & 107 && 3.9(-10) & 107  \\

\hline
\end{tabular}
\end{center}
\tablefoot{Results are given for each of the three warm-up timescales, using either high or low activation energies associated with CN addition to unsaturated hydrocarbons. 
X[i] = n[i]/n[H$_2$]. $A(B) = A \times 10^{B}$.
}
\end{table*}


\begin{table*}
\begin{center}
\caption{Ratios of branched-chain molecules with their straight-chain forms, for each model, as well as ratios between larger and smaller homologues. Observed ratios obtained by Belloche et al. (2014, 2016) are also shown.}
\label{tab-ratios}
\renewcommand{\arraystretch}{1.0}
\small
\begin{tabular}[t]{rcccccccccccccc}
\hline \hline
Molecular ratio && \multicolumn{3}{c}{Fast} && \multicolumn{3}{c}{Medium} && \multicolumn{3}{c}{Slow} && Observations \\
\cline{3-5} \cline{7-9} \cline{11-13} \\

 & & Low E$_A$ &&High E$_A$ && Low E$_A$ &&High E$_A$ && Low E$_A$ &&High E$_A$ \\
\hline

C$_3$H$_7$CN, \vspace{3mm}     $i$ / $n$              && 3.0     && 0.45    && 0.24       && 0.17     && 1.8      && 1.5    &&   $0.40 \pm 0.06$ $^a$  \\

C$_4$H$_9$CN,                           $i$ / $n$              && 0.84    && 1.5     && 0.61       && 0.60     && 1.9       && 2.2    &&  ---   \\
                                                  $s$ / $n$              && 4.3     && 4.1      && 2.4        && 2.6       && 1.9       && 1.7    &&  ---   \\
                                                   $t$ / $n$             && 0.042  && 0.027   && 0.015    && 0.013  && 0.10     && 0.085 &&  ---  \\
           \vspace{3mm} ($i$+$s$+$t$) / $n$             && 5.4     && 5.8      && 3.0        && 3.1      && 3.9       && 4.1    &&  ---  \\

C$_4$H$_{10}$, \vspace{3mm}    $i$ / $n$              && 0.13   && 0.13    && 0.055      && 0.054   && 0.36     && 0.36  &&  ---    \\

C$_5$H$_{12}$,                         $i$ / $n$              && 33      && 31       && 3.0         && 2.9       && 2.0       && 2.0    &&  ---  \\
                                                {\em neo} / $n$    && 0.28   && 0.27   && 0.11       && 0.10      && 0.48     && 0.48  &&  ---   \\
                \vspace{3mm} ({\em neo}+$i$) / $n$    && 34      && 32      && 3.2         && 3.0       && 2.6        && 2.6   &&  ---   \\

C$_2$H$_5$CN / CH$_3$CN                                   && 4.0     && 2.1     && 15         && 12         && 10       && 5.5    && 2.8 $^b$   \\
C$_2$H$_5$CN / C$_2$H$_3$CN                            && 9.2     && 7.5      && 6.5        && 6.2        && 4.7      && 3.8    && 15 $^b$  \\

$n$-C$_3$H$_7$CN / C$_2$H$_5$CN                                     && 0.024 && 0.14    && 0.13       && 0.17      && 0.054   && 0.67  && 0.029 $^{a,b}$  \\
C$_3$H$_7$CN ({\em total}) / C$_2$H$_5$CN                       && 0.098 && 0.21    && 0.16       && 0.20      && 0.15   && 0.16  && 0.041 $^{a,b}$  \\

$n$-C$_4$H$_9$CN / $n$-C$_3$H$_7$CN                              && 0.52   && 0.11    && 0.33       && 0.31      && 1.1      && 1.0    && $<0.59$ $^c$   \\
C$_4$H$_9$CN ({\em total}) / C$_3$H$_7$CN ({\em total})  && 0.82   && 0.52    && 1.1        && 1.1        && 1.8      && 2.1    && ---   \\

\hline
\end{tabular}
\end{center}
\tablefoot{$^a$ Belloche et al. (2014), $^b$ Belloche et al. (2016), $^c$ This work.
}

\end{table*}



\begin{figure*}
         {\includegraphics[width=0.5\hsize]{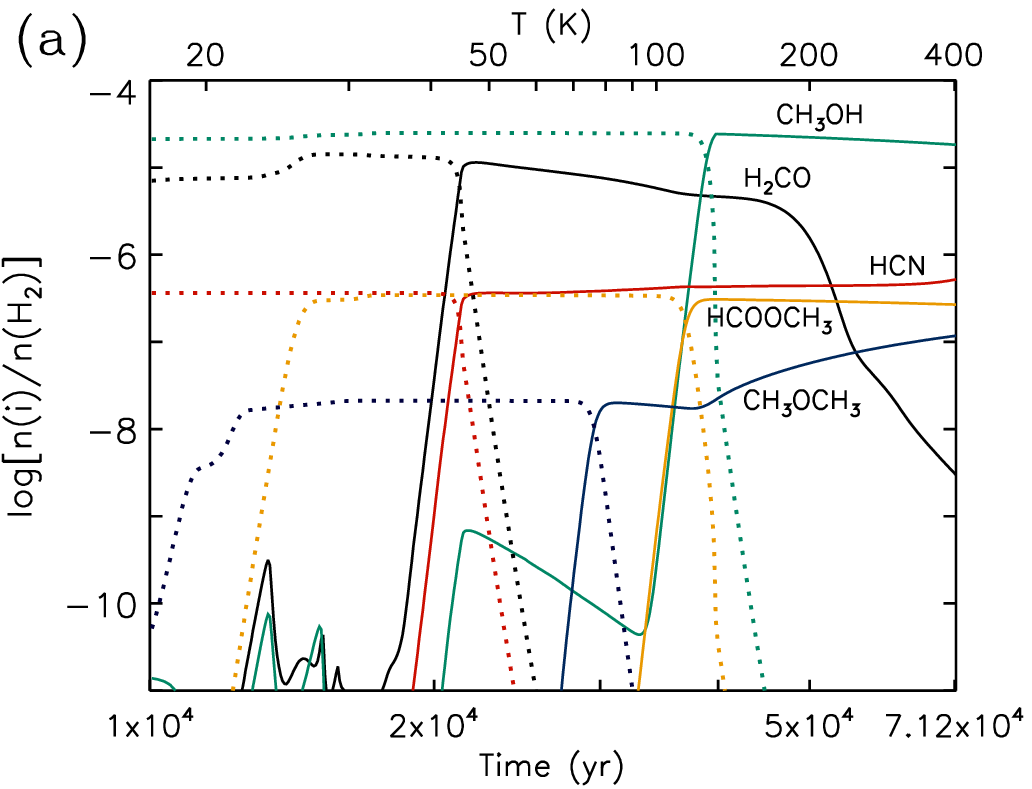}}
         {\includegraphics[width=0.5\hsize]{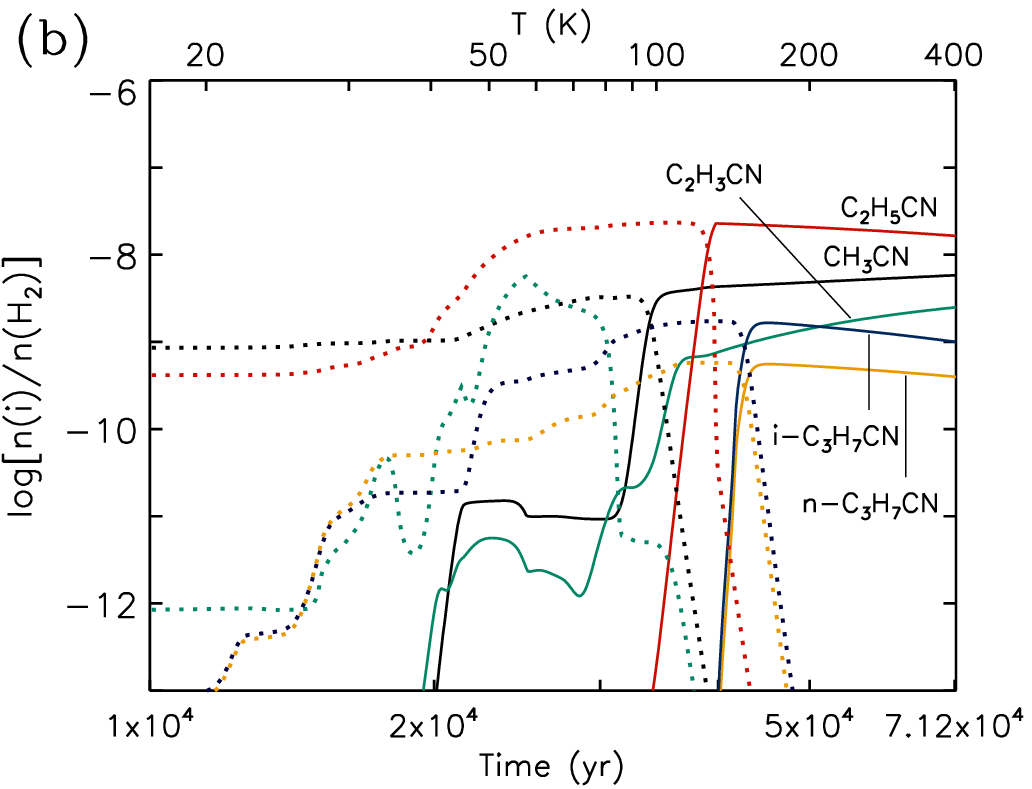}}
         {\includegraphics[width=0.5\hsize]{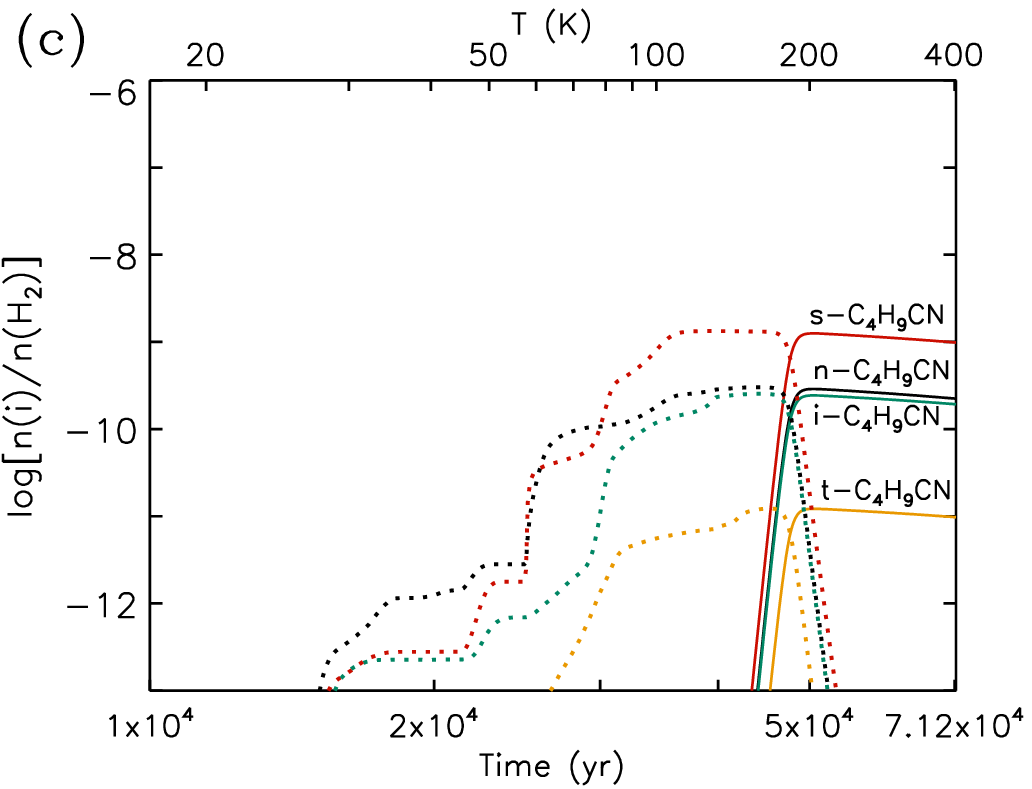}}
         {\includegraphics[width=0.5\hsize]{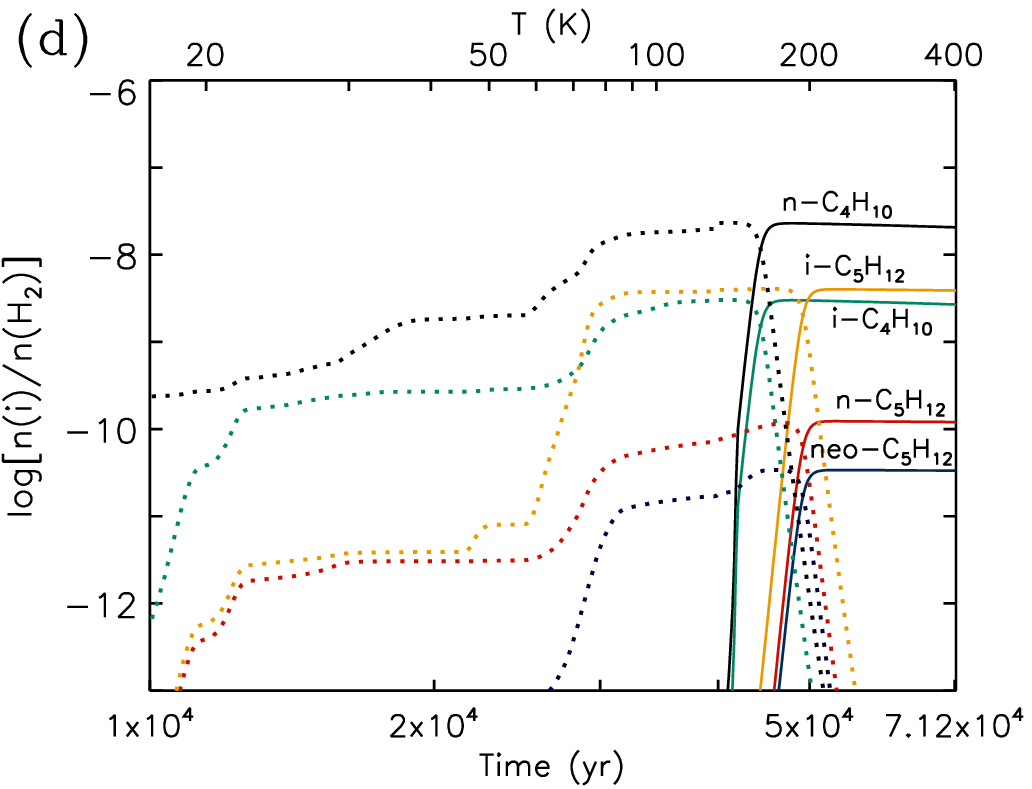}}
\caption{Calculated fractional abundances of nitriles, hydrocarbons and selected typical hot-core molecules for the {\em fast} warm-up model, assuming low (standard) barriers to CN-group addition to unsaturated hydrocarbons on grains. Solid lines indicate gas-phase species; dotted lines of the same colour represent the same species on the grains.}
\label{basic_fast}
\end{figure*}

\begin{figure*}
         {\includegraphics[width=0.5\hsize]{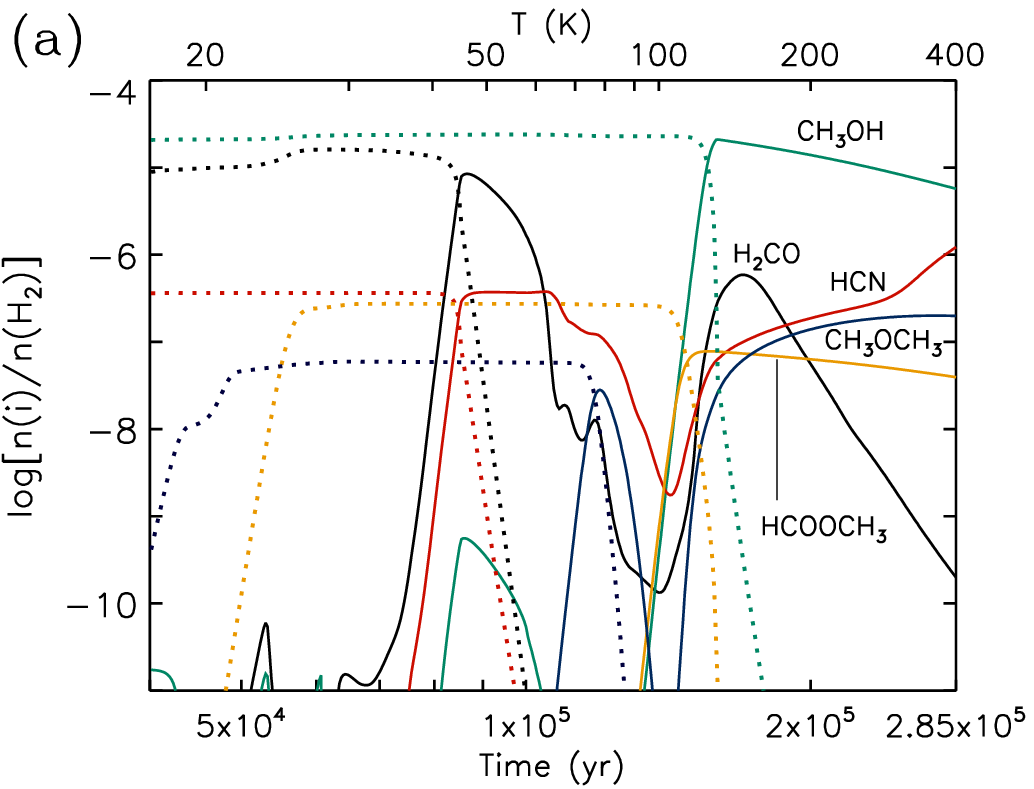}}
         {\includegraphics[width=0.5\hsize]{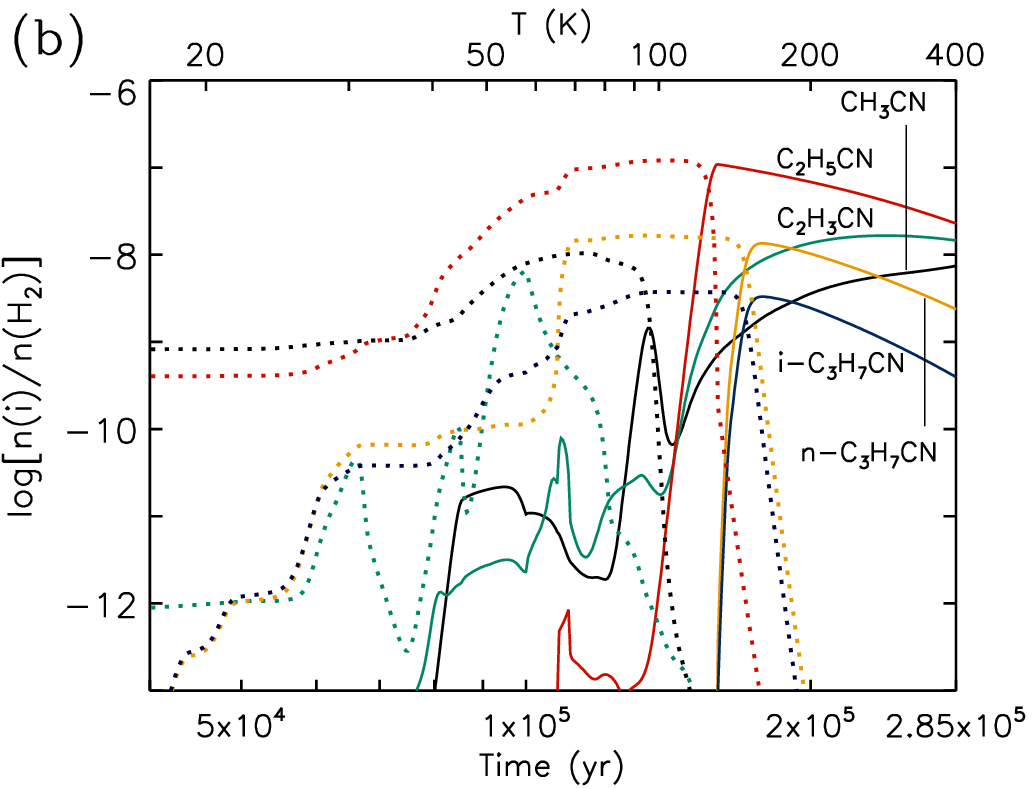}}
         {\includegraphics[width=0.5\hsize]{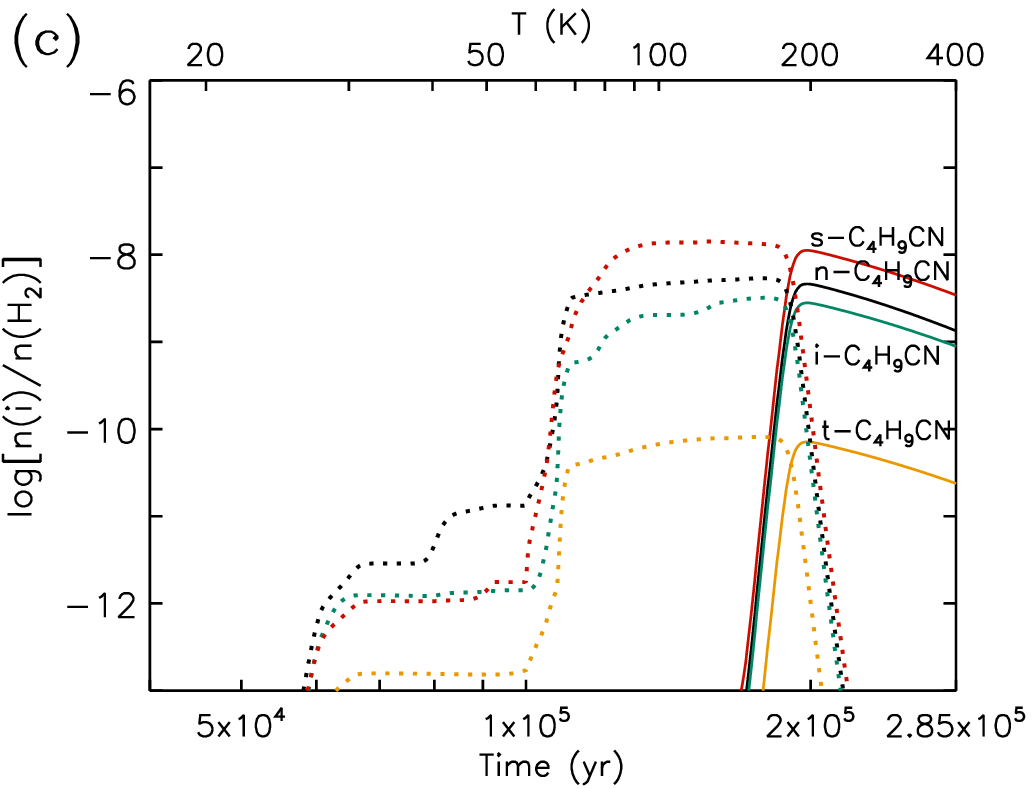}}
         {\includegraphics[width=0.5\hsize]{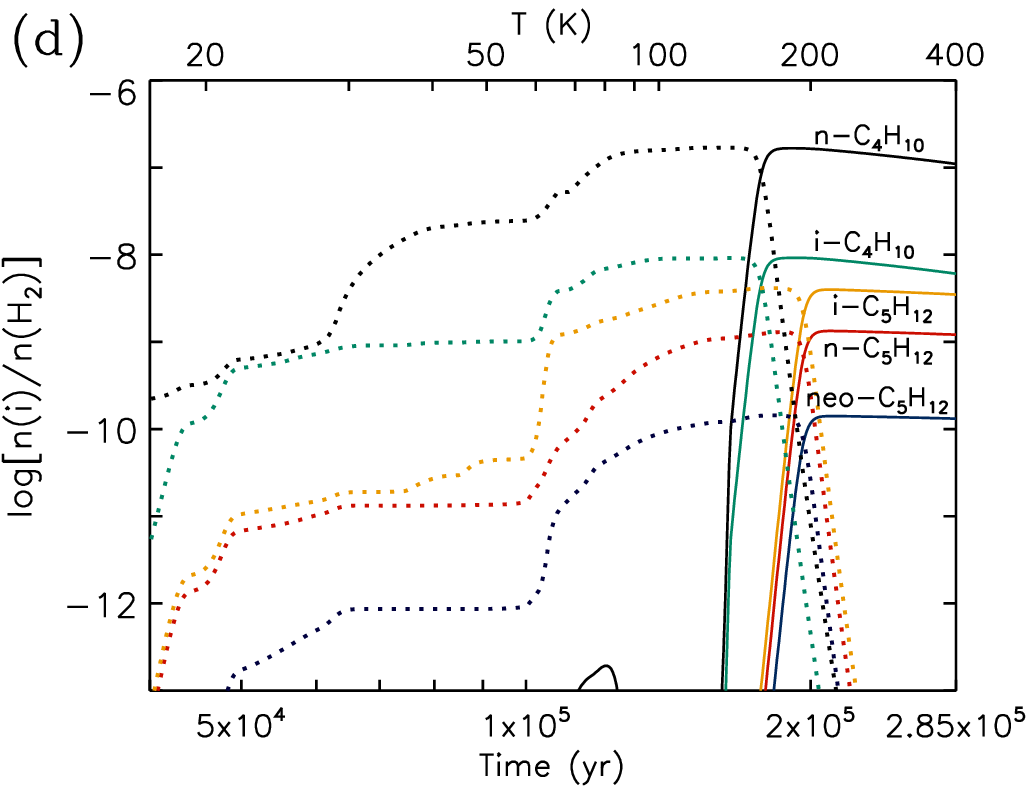}}
\caption{Same as for figure 2, for the {\em medium} warm-up model, assuming low (standard) barriers to CN-group addition to unsaturated hydrocarbons on grains.}
\label{basic_mid}
\end{figure*}

\begin{figure*}
         {\includegraphics[width=0.5\hsize]{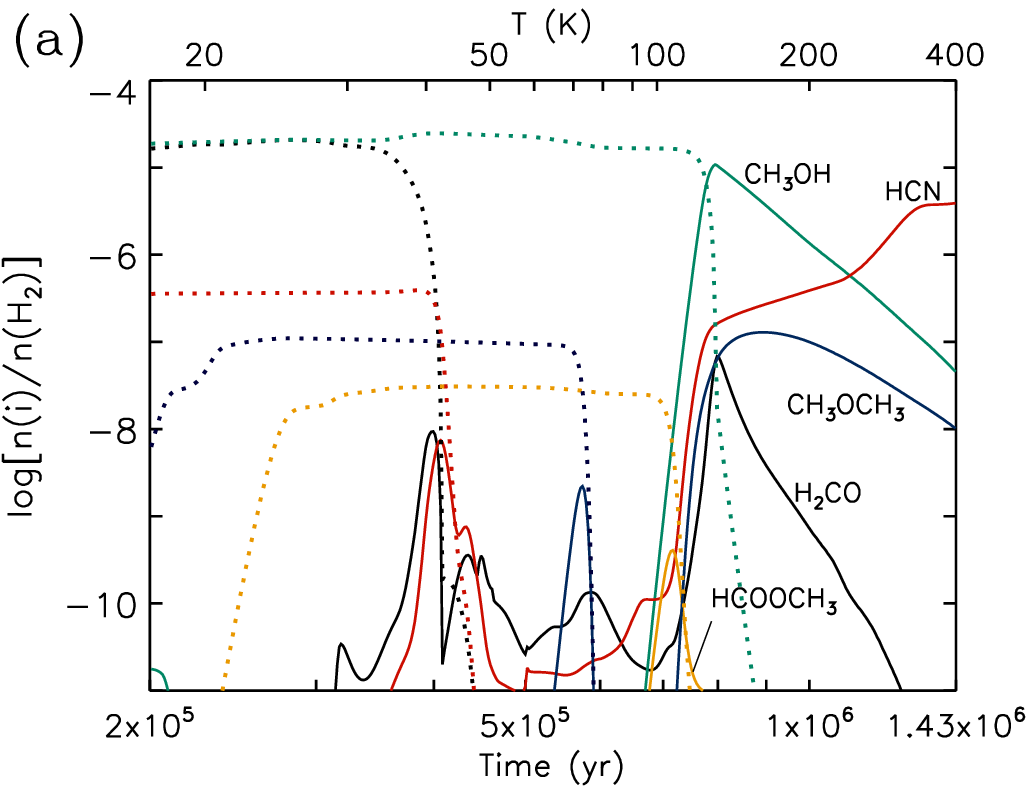}}
         {\includegraphics[width=0.5\hsize]{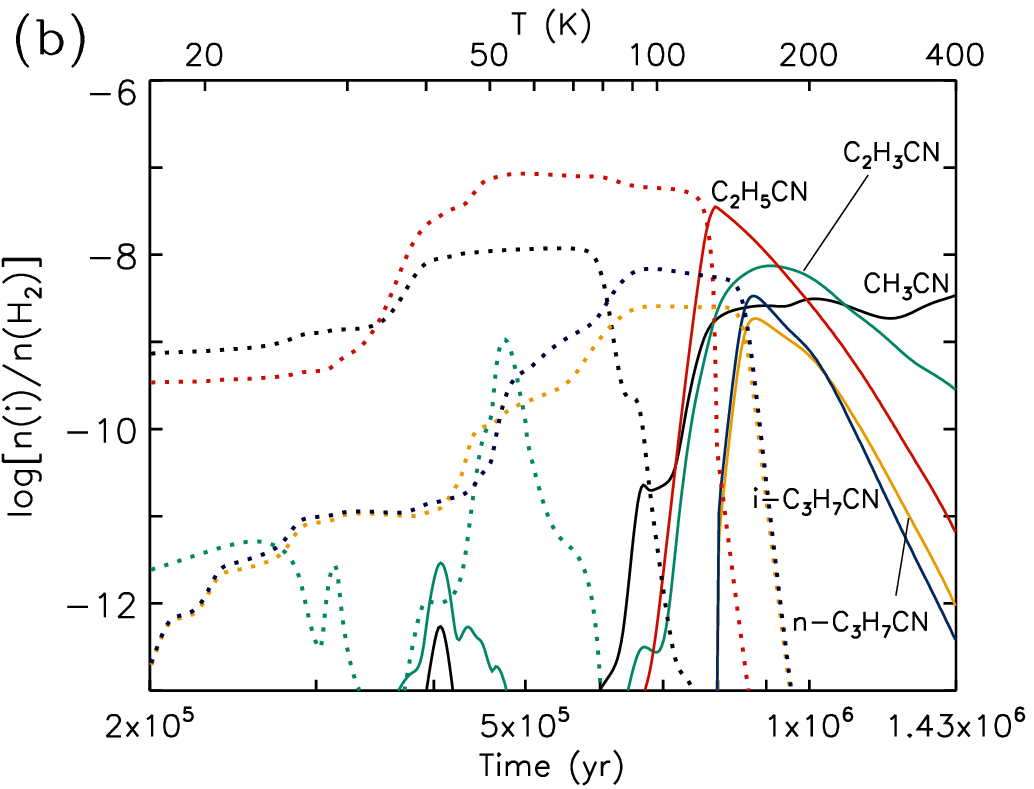}}
         {\includegraphics[width=0.5\hsize]{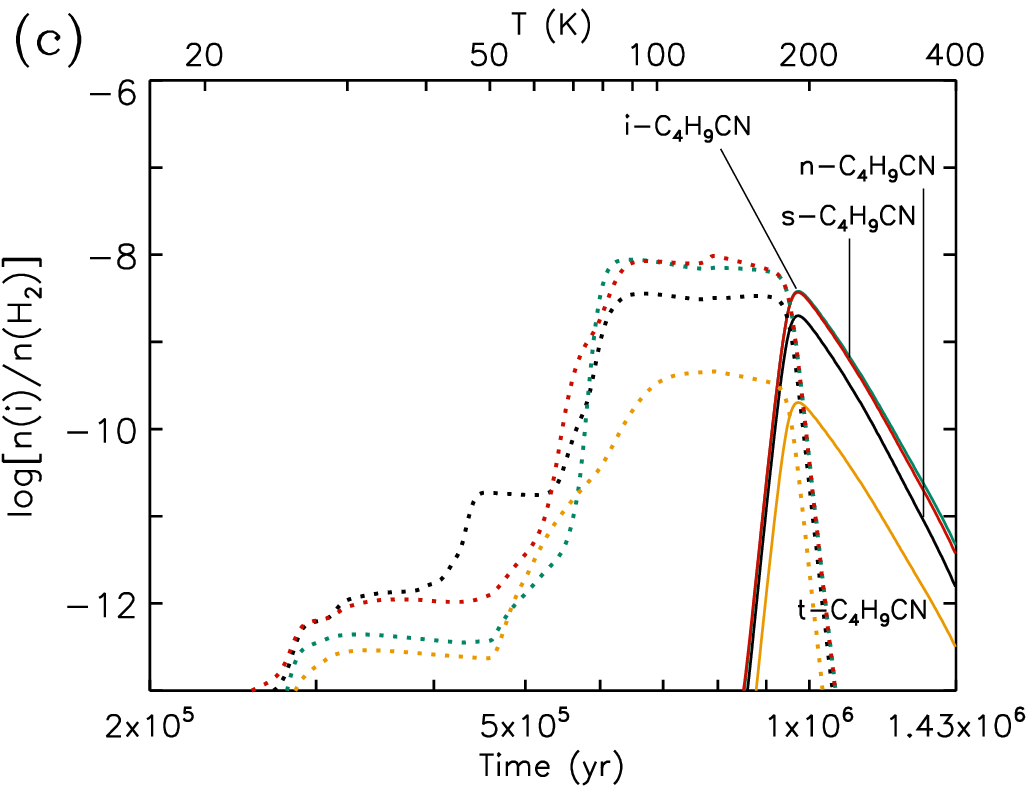}}
         {\includegraphics[width=0.5\hsize]{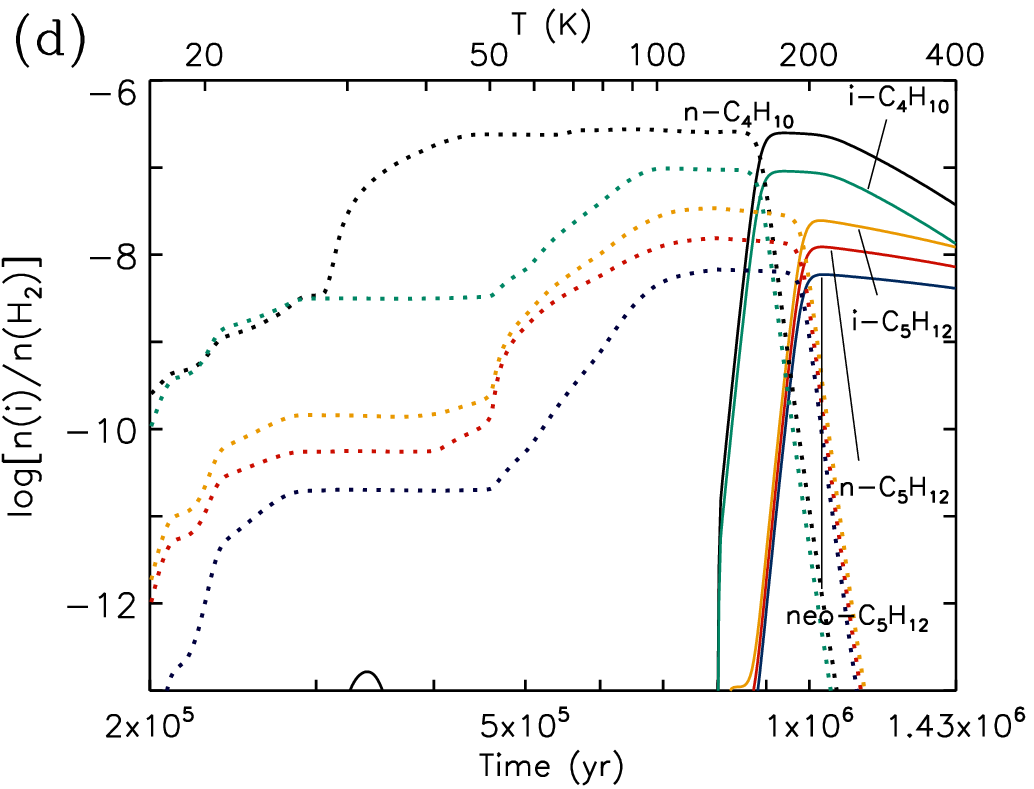}}
\caption{Same as for Figure 2, for the {\em slow} warm-up model, assuming low (standard) barriers to CN-group addition to unsaturated hydrocarbons on grains.}
\label{basic_slow}
\end{figure*}

\begin{figure*}
         {\includegraphics[width=0.5\hsize]{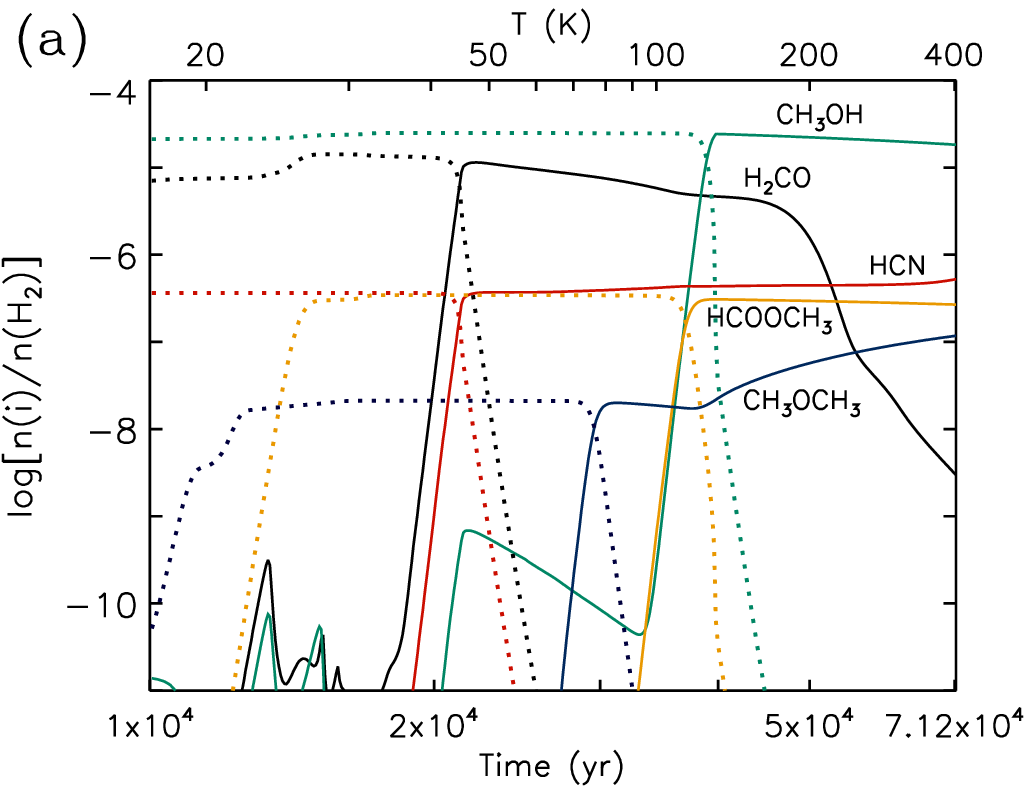}}
         {\includegraphics[width=0.5\hsize]{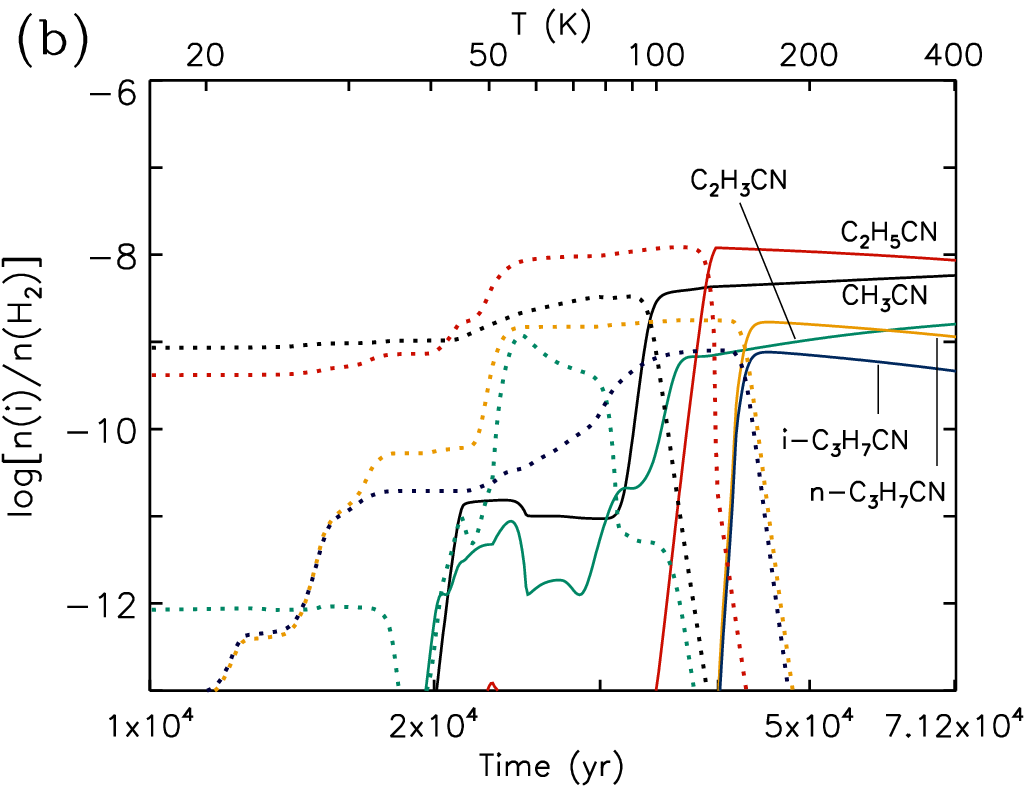}}
         {\includegraphics[width=0.5\hsize]{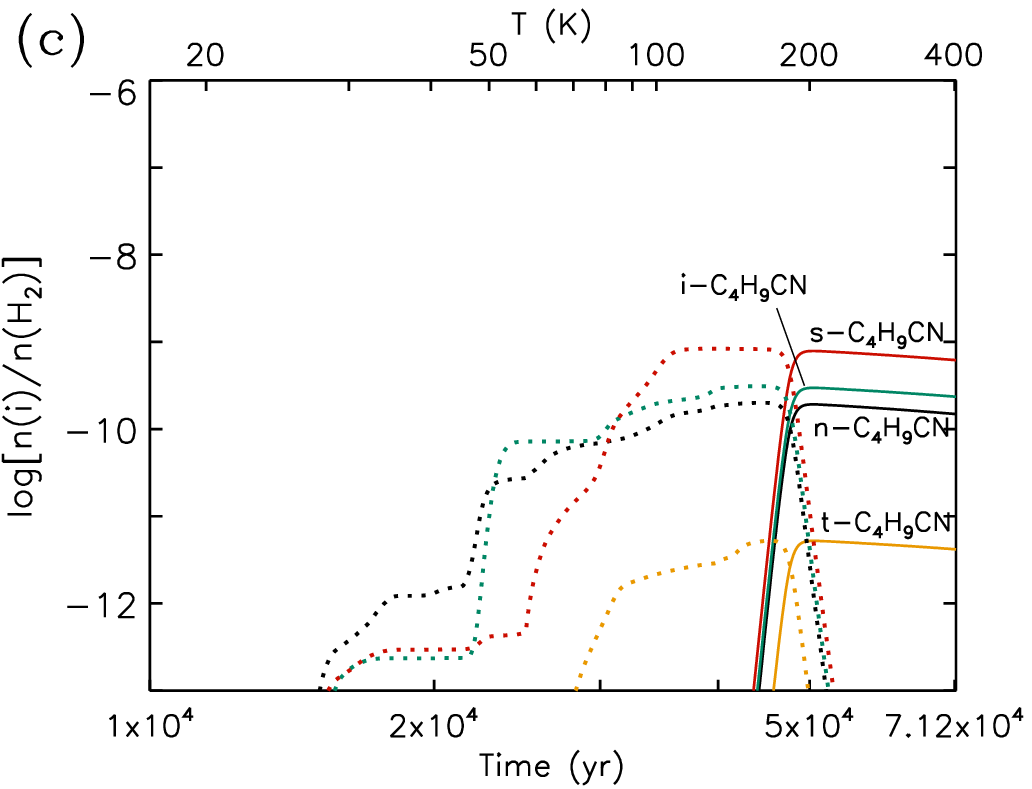}}
         {\includegraphics[width=0.5\hsize]{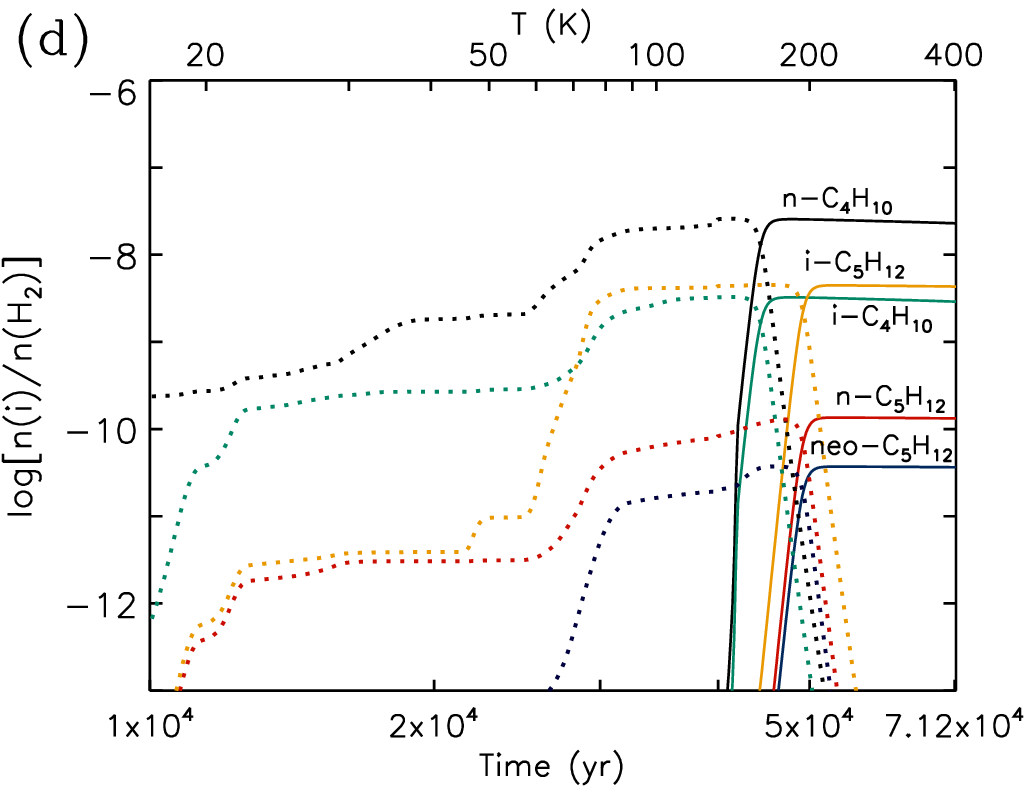}}
\caption{Same as for Figure 2, for the {\em fast} warm-up model, assuming {\em elevated} barriers to CN-group addition to unsaturated hydrocarbons on grains.}
\label{CN-barrs_fast}
\end{figure*}


\begin{appendix}
\section{Monte Carlo simulation of diffusive random walk to surface in the ice mantle}

In order to model accurately the rate at which molecules in the ice mantle reach the surface layer through isotropic thermal diffusion (assumed here to occur through a series of discrete ``swapping'' events), we need a description of the mean number of diffusion events required for a particle of arbitrary position to diffuse to the surface of an ice of any given thickness. This diffusion process is a three-dimensional random walk, but we are interested only in a particle's progress in one of those dimensions (i.e. depth), for which both the upper and lower bounds are explicitly defined by the thickness of the ice.

To determine the required relationship, we construct and apply a simple Monte Carlo random-walk diffusion model. This ice-mantle model consists of a cubic lattice of finite thickness in monolayers, $N_{\mathrm{thick}}$, with arbitrary size in the two lateral dimensions. For each cycle of the simulation, a single mobile particle is placed in a specific layer in the lattice, and the number of discrete diffusion events (or moves) required to reach the surface layer from that starting point is counted. Because we are interested only in the number of moves, no explicit rate calculations are required in this model. 

A single move upward or downward in the ice constitutes a move from one contiguous ice layer to another. The particle is allowed to move in any of the six directions within the lattice with equal probability, and all moves -- in any direction -- are added to the total count. If the particle reaches the lowest layer, movement in only the remaining five directions is allowed, again with equal probability. A random number generator is used to choose the direction of each move in the simulation. The simulation ends when the particle moves out of the top layer. 

The full simulation is repeated (with new random numbers) a minimum of 20,000 times per initial position (i.e. initial layer), with calculations carried out with a starting position at every one of the $N_{\mathrm{thick}}$ layers; for values of $N_{\mathrm{thick}} \leq 10$, the simulations are repeated 10 million times per initial layer. The data from these simulations provide the mean number of moves required to reach the surface from each starting layer, for a fixed ice thickness. The mean of this number over all layers then gives the average number of moves required to reach the surface from an arbitrary position within that ice mantle, which quantity we label $N_{\mathrm{move}}$. We then vary $N_{\mathrm{thick}}$ to test the overall dependence of this quantity on ice thickness.

\begin{figure}
\centerline{\resizebox{1.0\hsize}{!}{\includegraphics[angle=0]
{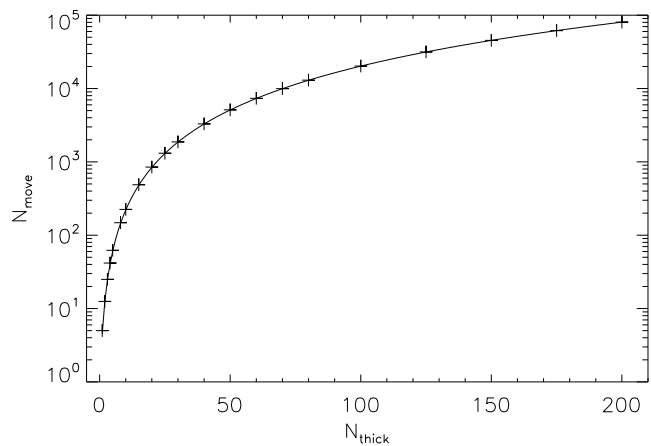}}}
\caption{Mean number of discrete diffusion events (moves) required for a particle of arbitrary position in an ice mantle of thickness $N_{\mathrm{thick}}$ to reach the surface. Crosses mark the results of the Monte Carlo simulations. The solid line denotes our fit to these data.}
\label{fit_fig}
\end{figure}

Figure \ref{fit_fig} shows the mean number of moves required to reach the surface for a range of ice thicknesses appropriate to interstellar grain-surface ices, as calculated using the Monte Carlo model (denoted by crosses). It is clear that the number of moves is not linear with the ice thickness. We find that the simple function
\begin{equation}
N_{\mathrm{move}} = 2 \, \left(N_{\mathrm{thick}} + \frac{1}{2} \right)^2
\end{equation}
\noindent provides an excellent fit to the data (shown as a solid line in Fig. \ref{fit_fig}). The fit matches all Monte Carlo values greater than $N_{\mathrm{thick}}$=1 to within 3~\% or less, with the fit improving steadily for $N_{\mathrm{thick}}$$>$4. At $N_{\mathrm{thick}}$=1, the Monte Carlo model produces an average of 5 moves, equal to the number of possible moves available to a particle at any moment when only one layer of ice exists in the mantle. Relationship (A.1) gives a value 4.5, so in practice we truncate the fit so that the adopted value of $N_{\mathrm{move}}$ may be no less than 5. The new fit is used with {\em MAGICKAL} in place of the $N_{\mathrm{move}}$ = $N_{\mathrm{thick}}$ relationship assumed by Garrod (2013). The new expression used in the present publication is shown in equation (1).

\end{appendix}


\begin{thebibliography}{}

\bibitem[Atkinson et al.(2001)]{} Atkinson, R., Baulch, D. L., Cox, R. A., Crowley, J. N., Hampson, R. F., Jr., Kerr, J. A., Rossi, M. J. \& Troe, J. 2001, {\em Summary of Evaluated Kinetic and Photochemical Data for Atmospheric Chemistry}, IUPAC Subcommittee on Gas Kinetic Data Evaluation for Atmospheric Chemistry -- Web Version

\bibitem[Baulch et al.(1992)]{Baulch1992} 
Baulch, D. L., Cobos, C. J., Cox, R. A., et al. 
1992, J. Phys. Chem. Ref. Data, 21, 411

\bibitem[Belloche et al.(2009)]{Belloche09} 
Belloche, A., Garrod, R.~T., M{\"u}ller, H.~S.~P., et al. 
2009, \aap, 499, 215

\bibitem[Belloche et al.(2014)]{Belloche14} 
Belloche, A., Garrod, R.~T., M{\"u}ller, H.~S.~P., \& Menten, K.~M. 
2014, Science, 345, 1584 

\bibitem[Belloche et al.(2016)]{Belloche16} 
Belloche, A., M{\"u}ller, H.~S.~P., Garrod, R.~T., \& Menten, K.~M. 
2016, \aap, 587, A91

\bibitem[Curran(2006)]{Curran2006} Curran, H. J. 2006, Int. J. Chem. Kinet., 38, 250

\bibitem[De Pree et al.(2015)]{DePree15} De Pree, C. G., Peters, T., Mac Low, M. M., et al. 2015, \apj, 815, 123

\bibitem[Gannon et al.(2007)]{Gannon2007} 
Gannon, K. L., Glowacki, D. R., Blitz, M. A., Hughes, K. J., Pilling, M. J. \& Seakins, P. W.
2007, J. Phys. Chem. A, 111, 6679 

\bibitem[Garrod(2008)]{Garrod08} Garrod, R. T. 2008, \aap, 491, 239

\bibitem[Garrod(2013)]{Garrod13} Garrod, R. T. 2013, \apj, 765, 60

\bibitem[Garrod et al.(2008)]{GWWH} Garrod, R. T., Widicus Weaver, S. L. \& Herbst, E. 2008, \apj, 682, 283

\bibitem[Garrod \& Pauly(2011)]{GP11} Garrod, R. T. \& Pauly, T. 2011, \apj, 735, 15

\bibitem[Gibb et al.(2000)]{Gibb00} Gibb, E. L., Whittet, D. C. B., Schutte, W. A., Boogert, A. C. A., Chiar, J. E., Ehrenfreund, P., Gerakines, P. A., Keane, J. V., Tielens, A. G. G. M., van Dishoeck, E. F., \& Kerkhof, O. 2000, \apj, 536, 347

\bibitem[Greiner(1970)]{} Greiner, N. R. 1970, J. Chem. Phys., 53, 1070

\bibitem[Guzm{\'a}n et al.(2015)]{} Guzm{\'a}n, A., Sanhueza, P., Contreras, Y., et al. 2015, \apj, 815, 130

\bibitem[Hasegawa \& Herbst(1993)]{hasegawa93} Hasegawa, T. I. \& Herbst, E. 1993, \mnras, 263, 589

\bibitem[Hennig \& Wagner(1995)]{} Hennig, G. \& Wagner, H. G. 1995, Ber. Bunsenges. Phys. Chem., 99, 863

\bibitem[Herbst \& Leung(1986)]{} Herbst, E \& Leung, C. M. 1986, \apj, 310, 378

\bibitem[Hong et al.(2016)]{Hong16} Hong, J., Mori, K., Hailey, C. J., et al. 2016, \apj, 825, 132

\bibitem[Hu et al.(1997)]{} Hu, W-P., Rossi, I., Corchado, J. C. \& Truhlar, D. G. 1997, J. Phys. Chem. A, 101, 6911

\bibitem[Kurylo \& Knable(1984)]{} Kurylo, M. J. \& Knable, G. L. 1984, J. Phys. Chem., 99, 3305


\bibitem[Mckee et al.(2007)]{} Mckee, K.W., Blitz, M.A., Cleary, P.A., Glowacki, D.R., Pilling, M.J., Seakins, P.W., Wang, L.M. 2007, J. Phys. Chem. A, 111, 4043 - 4055

\bibitem[Michael et al.(2005)]{Michael2005} 
Michael, J. V., Su, M. C., Sutherland, J. W., Harding, L. B. \& Wagner, A. F.
2005, Proc. Combust. Inst., 30, 965

\bibitem[M{\"u}ller et al. (2016)]{Mueller16}
M{\"u}ller, H. S. P., Belloche, A., Xu, L.-H., Lees, R. M., Garrod, R. T., Walters, A., van Wijngaarden, J., Lewen, F., Schlemmer, S. \& Menten, K. M.
2016, \aap, 587, A92

\bibitem[{\"O}berg et al.(2011)]{Oberg11} {\"O}berg, K. I., Boogert, A. C. A., Pontoppidan, K. M., van den Broek, S., van Dishoeck, E. F., Bottinelli, S., Blake, G. A. \& Evans, N. J., 2011 \apj, 740, 109

\bibitem[Ordu et al.(2012)]{Ordu12} Ordu, M.~H., M{\"u}ller, H.~S.~P., Walters, A., et al.\ 2012, \aap, 541, A121

\bibitem[Parker et al.(2004)]{} Parker, J. K., Payne, W. A., Cody, R. J. \& Stief, L. J. 2004, J. Phys. Chem. A., 108, 1938

\bibitem[Rolffs et al.(2011)]{Rolffs11} Rolffs, R., Schilke, P., Wyrowski, F., et al.\ 2011, \aap, 527, A68

\bibitem[Smith et al.(2004)]{} Smith, I. W. M., Herbst, E. \& Chang, Q. 2004, \mnras, 350, 323

\bibitem[Tsang(1987)]{} Tsang, W. 1987, J. Phys. Chem. Ref. Data, 16, 471

\bibitem[van der Tak et al.(2006)]{vdTak06} van der Tak, F. F. S., Belloche, A., Schilke, P., G{\"u}sten, R., Philipp, S., Comito, C., Bergman, P. \& Nyman, L.-\AA. 2006, \aap, 454, 99
	
\bibitem[Vigren et al.(2010)]{} Vigren, E., Hamberg, M., Zhaunerchyk, V., Kaminska, M., Thomas, R. D., Trippel, S., Wester, R., Zhang, M., Kashperka, I., af Ugglas, M., Semaniak, J., Larsson, M. \& Geppert, W. D. 2010, \apj, 722, 847


\end{thebibliography}
\end{document}